\documentclass[times,final]{elsarticle}

\usepackage{jasr}
\usepackage{framed,multirow}

\usepackage{amssymb}
\usepackage{latexsym}

\usepackage{amsmath}
\usepackage{threeparttable}
\usepackage{graphicx}

\usepackage[switch]{lineno}

\usepackage{url}
\usepackage{xcolor}
\definecolor{newcolor}{rgb}{.8,.349,.1}

\usepackage[citebordercolor=white]{hyperref}

\journal{Advances in Space Research}

\begin{document}

\verso{N. N. Shakhvorostova \textit{etal}}

\begin{frontmatter}

\title{Radioastron probing the fine structure of the flaring \texorpdfstring{H$_2$O}{H2O} maser \\ in star-forming region \texorpdfstring{G25.65+1.05}{G25}}%

\author[1]{N. N. \snm{Shakhvorostova}\corref{cor1}}
\cortext[cor1]{Corresponding author: 
  Tel.: +7-495-333-2301;  
  fax: +7-495-333-2378;}
  \ead{nadya.shakh@gmail.com}
\author[1]{A. V. \snm{Alakoz}}
  \ead{l-sha@yandex.ru}
\author[1]{I. E. \snm{Val'tts}}
 \ead{ivaltts@asc.rssi.ru}
\author[1]{I. D. \snm{Litovchenko}}
 \ead{ivanastronom@gmail.com}

\affiliation[1]{organization={P.N. Lebedev Physical Instutute of the Russian Academy of Sciences},
               addressline={53 Leninskiy prospekt},
               city={Moscow},
               postcode={119991},
              country={Russia}}


\received{?? May 2025}
\finalform{?? May 2025}
\accepted{?? May 2025}
\availableonline{?? May 2025}
\communicated{S. Sarkar}


\begin{abstract}
The paper describes a Space-VLBI observation of the 22 GHz H$_2$O masers in the massive star-forming region G25.65+1.05, using the 10-m space antenna of RadioAstron together with the ground-based VLBA array. Two observing epochs at the pre-flare and post-flare state of the maser source are presented. Leveraging the exceptional angular resolution provided by space-ground baselines along with the broad \texttt{UV} coverage from VLBA baselines, we gained a detailed perspective on the area associated with the maser flare events.
\end{abstract}


\end{frontmatter}

\nolinenumbers

\section{Introduction}
\label{intro}

Super-flares of maser emission is one of the most intriguing phenomena associated with dynamic processes within star-forming regions. During super-flares, the maser flux density greatly increases compared with its normal state and may highlight different events accompanying star formation process. It is currently believed that maser flares are triggered by either an episodic accretion burst in a massive young stellar object \citep{2017NatPh..13..276C} or an overlap of two maser cloudlets along the line of sight of the observer \citep{2005ApJ...634..459S}. The former scenario usually involves class II methanol masers as in the textbook case of G358.93-0.03 \citep[e.\,g.][]{Sugiyama19, 2019ApJ...881L..39B, Burns20}, while the second is associated exclusively with water masers \citep[e.\,g.][]{Liljestrom2000, 2014PASJ...66..106H, 2019ARep...63..814C} with the most representative example of Orion~KL \citep{2011ApJ...739L..59H}. Both types of maser flares are currently being in the main focus of the Maser Monitoring Organization (M2O\footnote{\url{https://www.masermonitoring.com/}}) established in order to make use of the great potential of maser flares as indicators of transient phenomena in massive young stellar objects (MYSO).

\begin{figure*}
\centering
\includegraphics[scale=0.35]{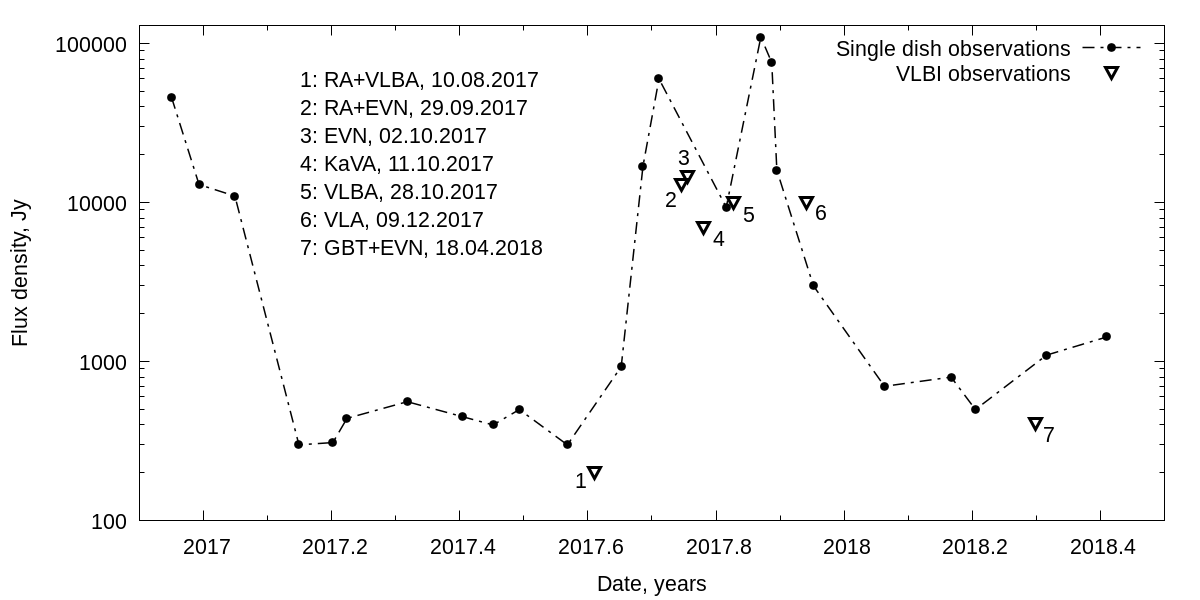}
\caption{The timeline of all conducted interferometric observations alongside with the single-dish data relevant to the period of super flares in 2016-2017 and VLBI sessions in 2017-2018. Circles indicate measurements obtained on 22-meter radio telescopes in Pushchino and Simeiz taken from \citep{2018ARep...62..213L, 2018ARep...62..584S, 2019ARep...63...49V, 2020ARep...64...15A, 2017ATel10853....1V, 2017ATel10728....1V, 2017ATel11042....1A}. Triangles with numbers indicate interferometric observations on different facilities including RadioAstron (RA), VLBA, EVN, KaVA and VLA. Dates of conducted VLBI sessions are shown in the label. References for the published data are as follows: 2~-- \citet{Bayandina2020}, 3~-- \citet{Burns2020}, 4 and 5~-- \citet{burns2018}, 6~-- \citet{Bayandina2019}. Observations 1 and 7 are presented in this work. 
}
\label{fig:timeline}
\end{figure*}

The massive star-forming region G25.65+1.05 (IRAS 18316-0602, RAFGL7009S) has been extensively studied in recent decade due to its spectacular super-flares of H$_2$O maser emission \citep{2018ARep...62..213L}, which placed this source among the brightest H$_2$O masers in our Galaxy, such as Orion~KL and W49N. In December 2016, a super-flare with flux density of 46\,000 Jy was detected in G25.65+1.05 during the regular monitoring of this source in Pushchino observatory \citep{2018ARep...62..213L}. This remarkable flare happened during the active operation of the RadioAstron space-ground interferometer \citep{Kardashev2017} and made G25.65+1.05 a subject of Cycle AO-5 RadioAstron maser observation program alongside with several other promising H$_2$O maser sources. In order to monitor the status of the sources and catch possible flares, the supporting single-dish monitoring of the proposed sources was organized in 2017 in Simeiz and Pushchino observatories at the two 22-m radio telescopes \citep{2018ARep...62..584S}. During this monitoring, the next super-flare in G25.65+1.05 was discovered in Simeiz observatory on September 7, 2017, where the flux density reached 17\,000 Jy \citep{2017ATel10853....1V, 2017ATel10728....1V, 2018ARep...62..584S, 2019ARep...63...49V}. This event started the extensive follow-up studies of the source using different facilities including single-dish telescopes \citep{2019ARep...63...49V, 2020ARep...64...15A} and interferometers RadioAstron \citep{Bayandina2020}, VLA \citep{Bayandina2019, Bayandina2023}, EVN \citep{Burns2020}, KaVA and VLBA~\citep{burns2018}, which revealed the structure of the source and the flare origin. The timeline of all conducted interferometric observations alongside with the single-dish data relevant to the period of super flares in 2016-2017 and VLBI sessions in 2017-2018 is presented in Figure~\ref{fig:timeline}. Flux densities are collected from the monitoring data obtained with 22-meter radio telescopes in Pushchino and Simeiz \citep{2018ARep...62..213L, 2018ARep...62..584S, 2019ARep...63...49V, 2020ARep...64...15A, 2017ATel10853....1V, 2017ATel10728....1V, 2017ATel11042....1A}. Fluxes corresponding to the interferometric data of observations 2, 3 and 6 (see Figure~\ref{fig:timeline}) are taken from the papers \citep{Bayandina2020, Burns2020, Bayandina2019, burns2018}. 

In the B-configuration VLA observations of \citet{Bayandina2019}, four centimeter continuum sources were detected and indicated as VLA 1-4. The 22 GHz H$_2$O maser flare was found to originate in the vicinity of the youngest continuum source in the region~-- VLA 1 \citep{Bayandina2019, Burns2020}, which is associated with H$_2$O maser emission only. Another continuum source VLA~2 is associated with both H$_2$O masers and 6.7~GHz methanol masers, however H$_2$O emission in its vicinity has not shown any significant flares until now (see discussion in Section~\ref{discussion}). Further investigation with higher angular resolution using VLA in A-configuration revealed that the continuum source VLA 1 is resolved into two separate sources VLA~1A and VLA~1B, and that the extraordinary flaring H$_2$O emission is associated particularly with the VLA~1A component \citep{Bayandina2023}. 
The VLA and EVN observations of the source during the H$_2$O maser flare revealed that the flare was due to the increased maser path length generated by the superposition of multiple maser emitting regions along the line of sight of the observer \citep{Bayandina2019, Burns2020}. RadioAstron's study of the H$_2$O maser super-flare in G25.65+1.05 region showed that the flare originated from a very compact maser region of a size of $\sim$25~$\mu$as (0.05 AU or $\sim$5 solar diameters at the adopted distance to the source of 2.08 kpc) and brightness temperature of $\sim$3$\times$10$^{16}$~K \citep{Bayandina2020}. RadioAstron, EVN and VLA sessions are indicated by numbers 2, 3 and 6 in the Figure~\ref{fig:timeline}, respectively.

There were three RadioAstron observing sessions dedicated to G25.65+1.05. The first one was conducted on 10 August 2017 in frames of a joint proposal for RadioAstron and VLBA facility while the source exhibited low brightness of $\sim$200~Jy, indicating a quiescent state (this paper). Since no interferometric observations preceding the super-flare have been published so far, the present study fills this gap. The second observation was carried out on 29 September 2017 during the ongoing flare detected on 7 September 2017, and apart from the 10-meter Space Radio Telescope (SRT) involved the 32-meter Torun radio telescope (Poland) and 26-meter Hartebeesthoek radio telescope (Republic of South Africa) \citep{Bayandina2020}. The third RadioAstron session was conducted on 18 April 2018 (this paper), in 7.5 months post-flare, when the source was again in relatively stable state according to the work \citep{2020ARep...64...15A}. The last observation involved the SRT antenna and several ground telescopes (see Section \ref{observations}), but unfortunately the SRT data were lost by technical reasons. Nevertheless, the ground-based data of this final session are of great importance because they were obtained in the period after a powerful outburst of G25.65+1.05 and allow us to trace the full evolution of the source spatial structure during its changing state. By studying the spatial distribution of H$_2$O masers both pre-outburst and well into the subsequent period, we aim to enrich our understanding of the source structure and trace its spatial evolution across a broader temporal interval than previously achieved. The other goal is to reveal the existence of possible super-compact maser spots related to the source and their origin using very high angular resolution supplied by Space-VLBI technique.


\section{Observations and data reduction}
\label{observations}

\begin{figure}
\centering
\begin{minipage}{0.49\linewidth}
\includegraphics[angle=270, scale=0.4]{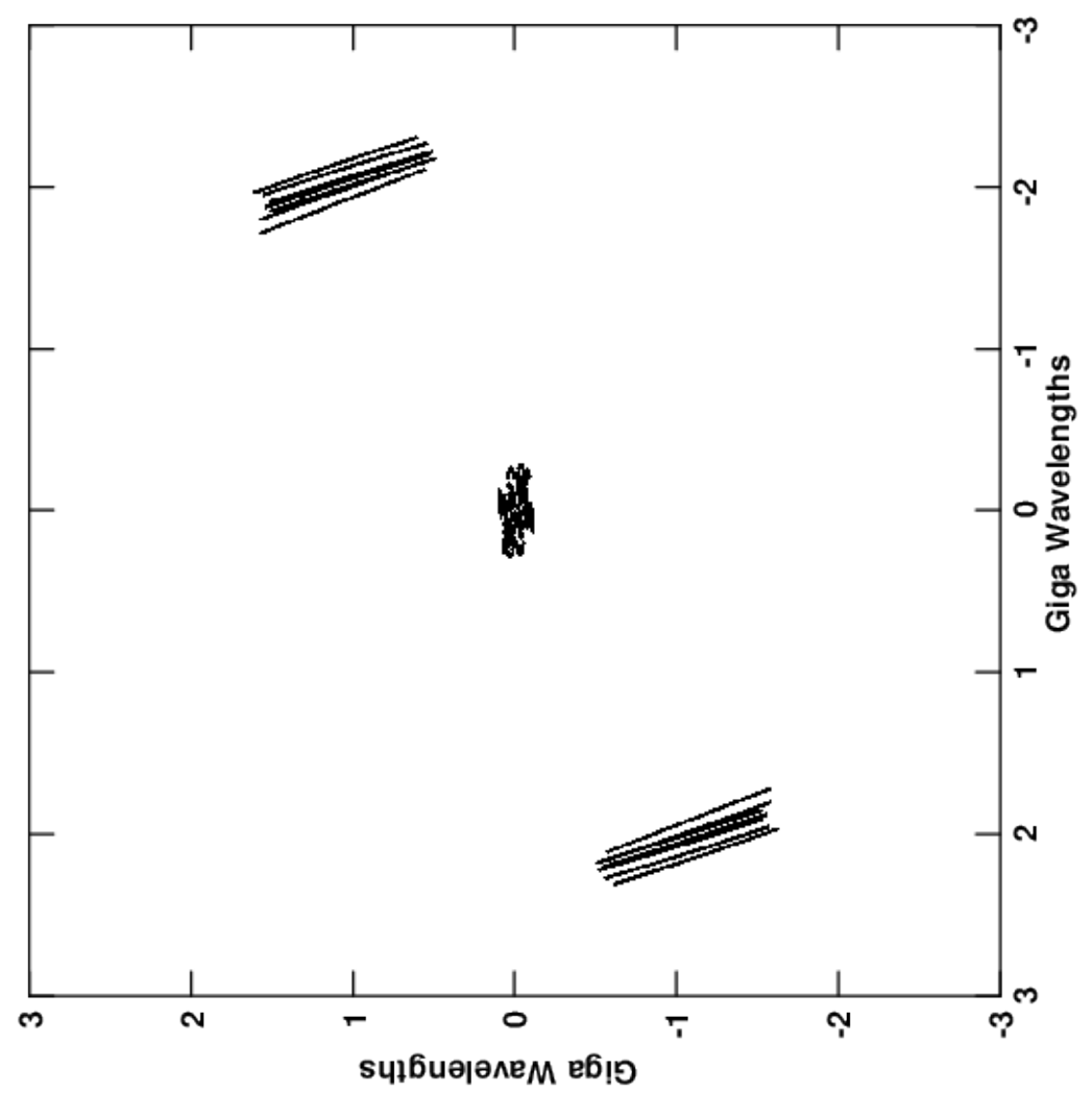}
\end{minipage}
\hfill
\begin{minipage}{0.49\linewidth}
\includegraphics[angle=270, scale=0.4]{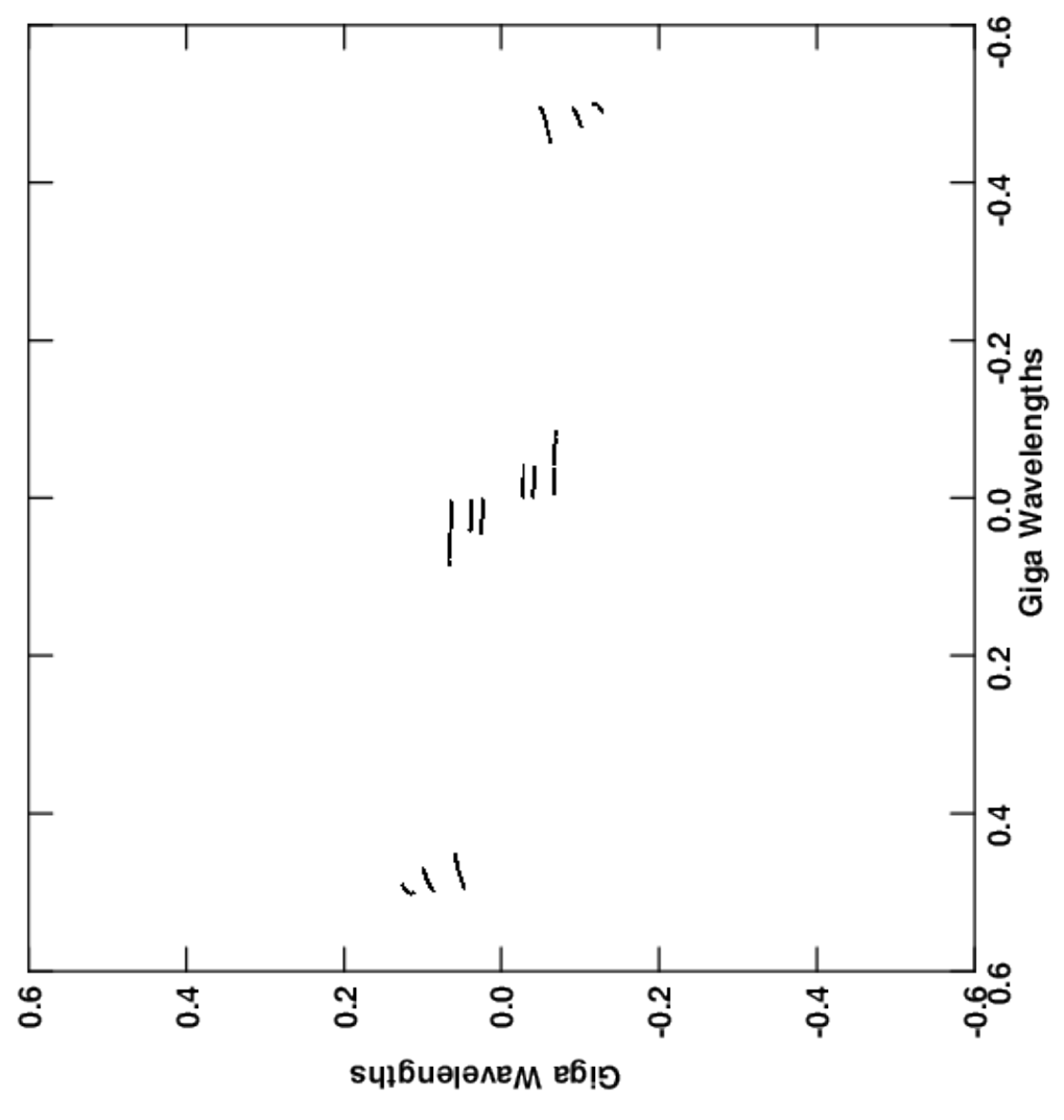}
\end{minipage}
\caption{The \texttt{UV}-plane coverage obtained in the RadioAstron SRT$-$VLBA observation on August 10, 2017 (Epoch~I, project code \texttt{rags31a}, left panel) and in the observing session conducted on April 18, 2018 using four ground-based radio telescopes without space antenna (Epoch~II, project code \texttt{rags31b}, right panel).}
\label{fig:UV}
\end{figure}

The first RadioAstron observation, conducted with the 10-meter space radio telescope and Very Long Baseline Array (VLBA), was performed on 10 August 2017 (project codes \texttt{rags31a} (RadioAstron) / \texttt{BK208} (VLBA)). The observation with ground array lasted six hours (01:00$-$07:00 UTC), while the space telescope joined the observation for the last two hours (05:00$-$07:00 UTC). The core array of eight VLBA telescopes (without St. Croix and Mauna Kea) participated in the observation. The obtained \texttt{UV}-coverage is presented in Figure \ref{fig:UV} (left panel): the central part of the plot up to $\sim$0.3 Giga wavelength corresponds to the ground-ground VLBA baselines, and the distant part at $\sim$2.5 Giga wavelength is obtained with the space-ground baselines. The space-ground baseline achieved $\sim$2.7 Earth Diameters (ED) which corresponds to the spatial resolution of $\sim$80 $\mu$as. 

The last observation of G25.65+1.05 in frames of RadioAstron program was performed on 18 April 2018 (project code \texttt{rags31b}) and involved ground-based 100-meter Green Bank telescope (GBT, USA) and 32-meter antennas in Svetloe (SV, Russia), Medicina (MC, Italy) and Torun (TR, Poland). The observation lasted almost three hours (03:15$-$06:05 UTC), but the Green Bank telescope joined it only for the last hour (05:00$-$06:05 UTC). Unfortunately, RadioAstron data were lost due to technical reasons, and only ground-based data were used in analysis. The obtained \texttt{UV}-coverage is presented in Figure \ref{fig:UV} (right panel) and the baselines range from 25 to 515 Mega wavelength. The maximum projected baseline 515 M$\lambda$ corresponds to the spatial resolution of $\sim$400 $\mu$as.

The H$_2$O maser 6$_{\rm 1,6}-5_{\rm 2,3}$ transition rest frequency is 22235.08 MHz. In both experiments, the center of the band was set at 22228.0 MHz, and the bandwidth was 32 MHz (two sub-bands of 16 MHz each). Both sub-bands contained 2048 spectral channels, which corresponds to a spectral resolution of 7.8125 kHz. The pointing coordinates of the target source G25.65+1.05 were set to $\alpha=18^{\rm h}$34$^{\rm m}$20.96$^{\rm s}$~(J\,2000) and $\delta=-05^{\circ}59^{\prime}42.15^{\prime\prime}$~(J\,2000) in \texttt{rags31a} observation and to $\alpha=18^{\rm h}$34$^{\rm m}$20.90$^{\rm s}$~(J\,2000) and $\delta=-05^{\circ}59^{\prime}41.46^{\prime\prime}$~(J\,2000) in \texttt{rags31b} observation. 

For calibration purposes, the following two fringe-finder sources were observed in the \texttt{rags31a} session: 3C~273 (J1229+0203) and 4C~09.57 (J1751+0939). Also, the phase reference calibrator J1821-0502 was observed at the angular separation from the target of 3.41$^{\circ}$. Due to technical limitations, the 10-meter space telescope was unable to handle fast-switching between the target and phase reference source. Only the fringe-finder source 4C~09.57 (1749+096) and the target were observed at the space-ground baselines. The VLBA part of the observation was conducted in the phase-reference mode, however the phase calibrator was not detected on most baselines and was not used in further analysis. In the \texttt{rags31b} observation, the following two calibration sources were observed: 3C~345 (J1642+3948) and 4C~09.57 (J1751+0939). The main parameters of the observations and sources are summarized in Tables~\ref{tab:obs} and \ref{tab:sources}.

\begin{table*}
\centering
\caption{Observation parameters}
\begin{tabular}{cccccccc}
\hline\hline
Observing & Epoch & Time & Baselines & Beam & Antennas & Resolution & Velocity \\
 code & & hours &  M$\lambda$ & mas$\times$mas, ($^{\circ}$) & & km/s (kHz) & km/s \\
 \hline
\texttt{rags31a}, & 10.08.2017 & 6  & 2170$-$2550 & 0.36$\times$0.11, $-$12.27 & SRT+VLBA  & 0.12 (7.8)  & $-$121 to +93\\
\texttt{BK208}    &            &    &  10$-$288 & 1.42$\times$0.51, $-$02.54   & VLBA &  &  \\
\hline
\texttt{rags31b} & 18.04.2018 & 3 & 25$-$515 & 2.10$\times$0.23, $-$10.80 & GBT, SV, MC, TR & 0.12 (7.8) & $-$77 to +138 \\
\hline\hline
\end{tabular}
\label{tab:obs}
\end{table*}

\begin{table*}
    \centering
    \caption{Observed sources\\   }
    \begin{tabular}{c|c|r|c|c}
    \hline\hline
   Source &    R.A.(J2000) &      DEC(J2000) & Observing & Comments\\
          &    hh:mm:ss.ssss   & dd:mm:ss.sss &  session & \\
    \hline
   G25.65+1.05 &  18:34:20.9600 & $-$05:59:42.150 & rags31a & target maser source \\
   1226+023 (J1229+0203) & 12:29:06.6997 & 02:03:08.598   & rags31a & fringe-finder source  \\
   1749+096 (J1751+0939) & 17:51:32.8186  & 09:39:00.728 & rags31a & delay calibrator \\
   1818-050 (J1821-0502) & 18:21:11.8095 & $-$05:02:20.086 & rags31a & phase calibrator \\
   \hline
   G25.65+1.05 &  18:34:20.9000 & $-$05:59:41.460 & rags31b & target maser source \\
   1641+399 (J1642+3948) & 16:42:58.8100 & 39:48:36.994 & rags31b & fringe-finder source \\
   1749+096 (J1751+0939) & 17:51:32.8186  & 09:39:00.728 & rags31b & delay calibrator \\
   \hline\hline
    \end{tabular}
    \label{tab:sources}
\end{table*}

The software correlator of the Astro-Space Center of the P.N.~Lebedev Physical Institute was used for the data correlation \citep{Likhachev2017}. The orbital parameters of the space antenna, required for the data correlation, were provided with the accuracy of 500 meters in position and 0.02 m/s in velocity \citep{Kardashev2013}. 
The phase center coordinates used for the final \texttt{rags31a} data correlation were set to $\alpha=18^{\rm h}$34$^{\rm m}$20.90$^{\rm s}$ and $\delta=-05^{\circ}59^{\prime}41.56^{\prime\prime}$~(J\,2000). It is different from the pointing coordinates (see Table~\ref{tab:sources}) since the recent VLA and EVN observations provided more precise location of the H$_2$O masers in the region \citep{Bayandina2019, Burns2020}. The calibration quasar 1226+023 was used to detect fringes and establish the initial values of group delay and phase delay rate. The phase center in the \texttt{rags31b} session was set to the pointing coordinates $\alpha=18^{\rm h}$34$^{\rm m}$20.90$^{\rm s}$~(J\,2000) and $\delta=-05^{\circ}59^{\prime}41.46^{\prime\prime}$~(J\,2000), and the calibration quasar 1641+399 was used for initial fringe search. 

The delay model employed in the correlation process took into account signal propagation delays through the standard atmosphere. In these experiments, we did not have telescope logs containing meteorological data. To calculate the atmospheric parameters, the U.S. Standard Atmosphere 1976 model was used\footnote{\url{https://www.pdas.com/atmos.html}}. The atmospheric parameters (air density, pressure and temperature) were approximated by polynomials, facilitating regularization and interpolation for input into the delay model. Zenith delay calculations were performed, comprising both hydrostatic (dry) and wet components. The last ones were computed using formulas derived from the work \citep{Saastamoinen1972}, while the hydrostatic delay constants were adopted from \citep{Davis1985}. Wet delay involved calculating the saturation vapor pressure using Murray’s formula \citep{Murray1967}. Finally, the zenith delay was transformed into an atmospheric delay along the source direction using Niell’s mapping functions \citep{Niell1996}.

The post-correlation data reduction was performed using the software package Astronomical Image Processing System (AIPS)\footnote{\url{http://www.aips.nrao.edu}} and VLBI processing software package PIMA\footnote{\url{https://astrogeo.org/pima}}. AIPS tasks \texttt{CVEL}, \texttt{FRING}, \texttt{ANTAB}, \texttt{ACCOR}, \texttt{APCAL}, \texttt{BPASS}, \texttt{ACFIT}, \texttt{CALIB} were used for the data calibration including fringe-fitting and amplitude calibration. Imaging was performed with AIPS task \texttt{IMAGR}. Package PIMA was used mainly for refinement of residual delay and fringe rate for the space antenna.


\section{Results}
\label{results}

\begin{figure*}
\centering
\includegraphics[scale=0.5]{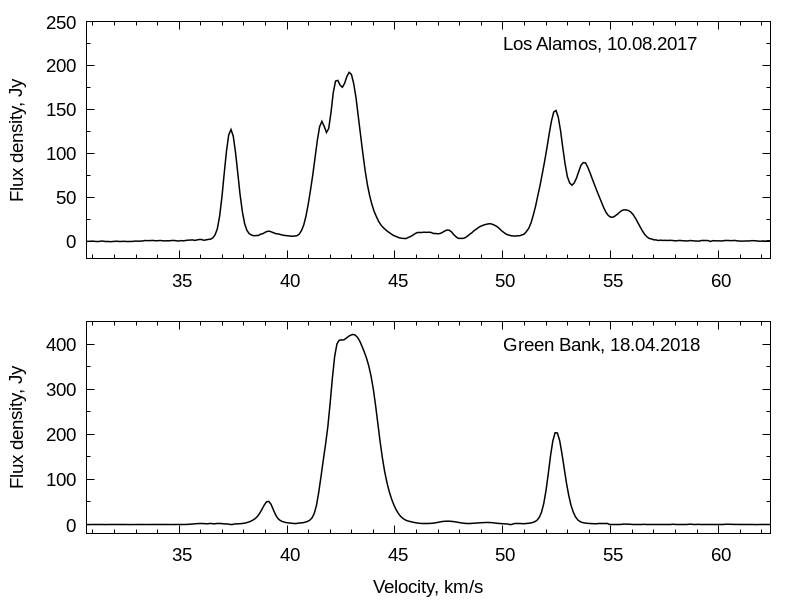}
\caption{The autocorrelation spectra of the 22 GHz H$_2$O maser in the high-mass star-forming region G25.65+1.05 obtained during two epochs: 1) on August 10, 2017 with 25-m Los Alamos radio telescope of the VLBA array (top panel) and 2) on April 18, 2018 with 100-m Green Bank radio telescope (bottom panel).}
\label{fig:AC}
\end{figure*}

Two observing epochs were conducted, as described above in Section \ref{observations} and Table \ref{tab:obs}. At the Epoch~I of the RadioAstron observation \texttt{rags31a} at 10 August 2017, the H$_2$O maser in G25.65+1.05 had a moderate flux density about $\sim$200~Jy (see Figure \ref{fig:AC}, top panel) and showed low activity. However, shortly after this observation the flux density of the source increased and reached 65~kJy by September 16, 2017 \citep{2019ARep...63...49V}. Therefore, the observational epoch of the \texttt{rags31a} experiment can be classified as a 'pre-flare' phase. As seen from Figure \ref{fig:AC} (top panel), the G25.65+1.05 spectrum exhibited a complex, multi-peak structure with numerous overlapping spectral features, which can be categorized into three groups: the blue spectral features at V$_{\rm LSR}$ $\sim$37-39 km s$^{-1}$, the central features (which include the flaring feature) centered around V$_{\rm LSR}$ $\sim$42 km s$^{-1}$, and the red spectral features located near V$_{\rm LSR}$ $\sim$53 km s$^{-1}$. Further we refer to these spectral groups as blue, bursting and red H$_2$O maser features.

The second epoch of observations \texttt{rags31b} were carried out after eight months, on 18 April 2018, during the post-flare of the maser G25.65+1.05. The autocorrelation spectrum obtained on Green Bank 100-m telescope is shown at the bottom panel in Figure~\ref{fig:AC}. The source also had a moderate flux density of 420~Jy, which is twice higher compared with the pre-flare flux measured in \texttt{rags31a} experiment. However the two spectra are quite different. Blue feature at 37.4~km/s disappeared and a faint feature at 39.2~km/s increased almost five times from 11 to 51 Jy, 
bursting feature two times increased, and the red features in range +51\,...\,+57~km/s turned to be a single peaked instead of three-peaked. 


\subsection{Red features and a compact \texorpdfstring{H$_2$O}{H2O} maser}

\begin{figure*}
\centering
\begin{minipage}[h]{0.49\textwidth}
\center{\includegraphics[scale=0.32, angle=270]{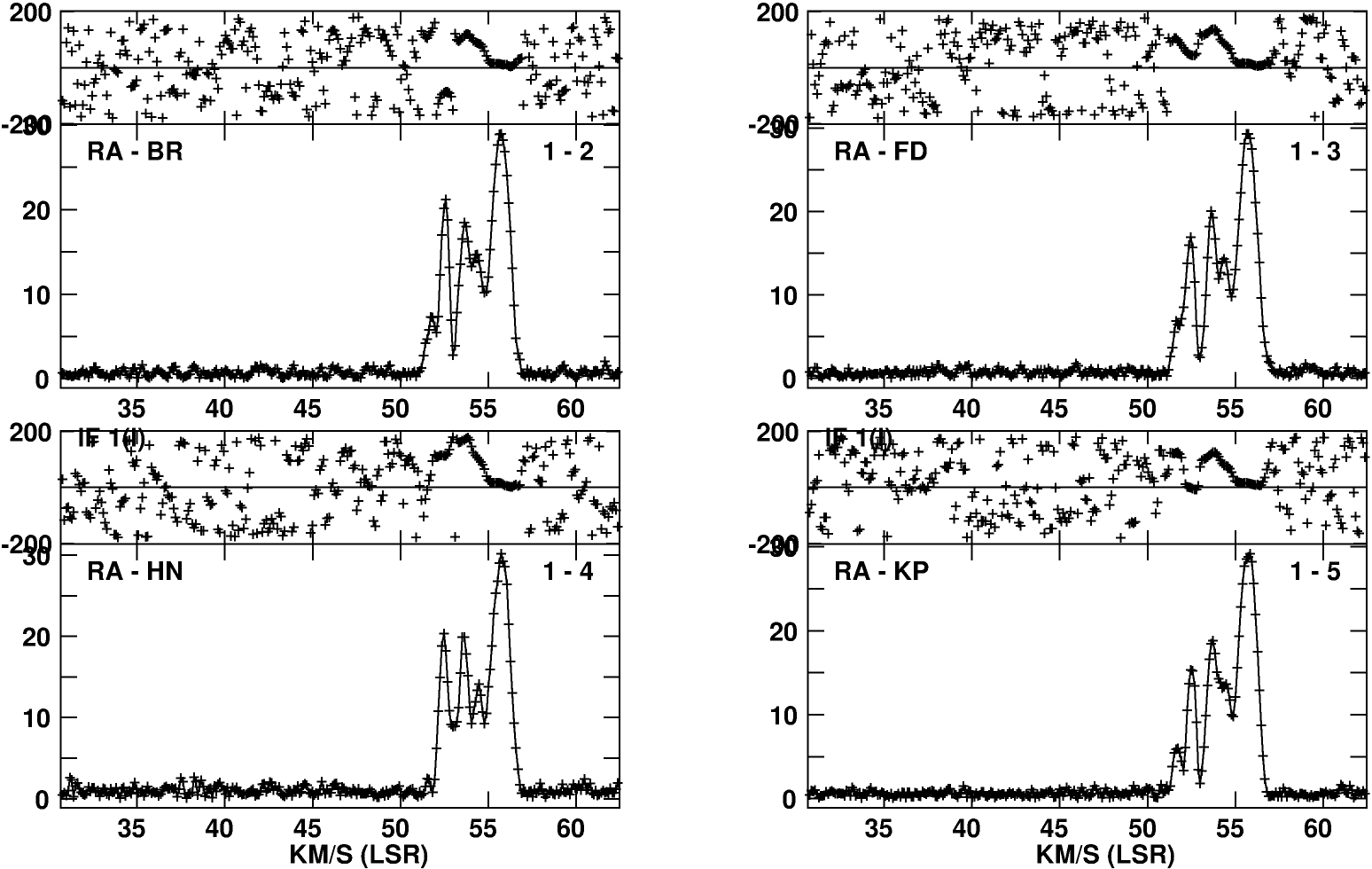}}
\end{minipage}
\hfill
\begin{minipage}[h]{0.49\textwidth}
\center{\includegraphics[scale=0.32, angle=270]{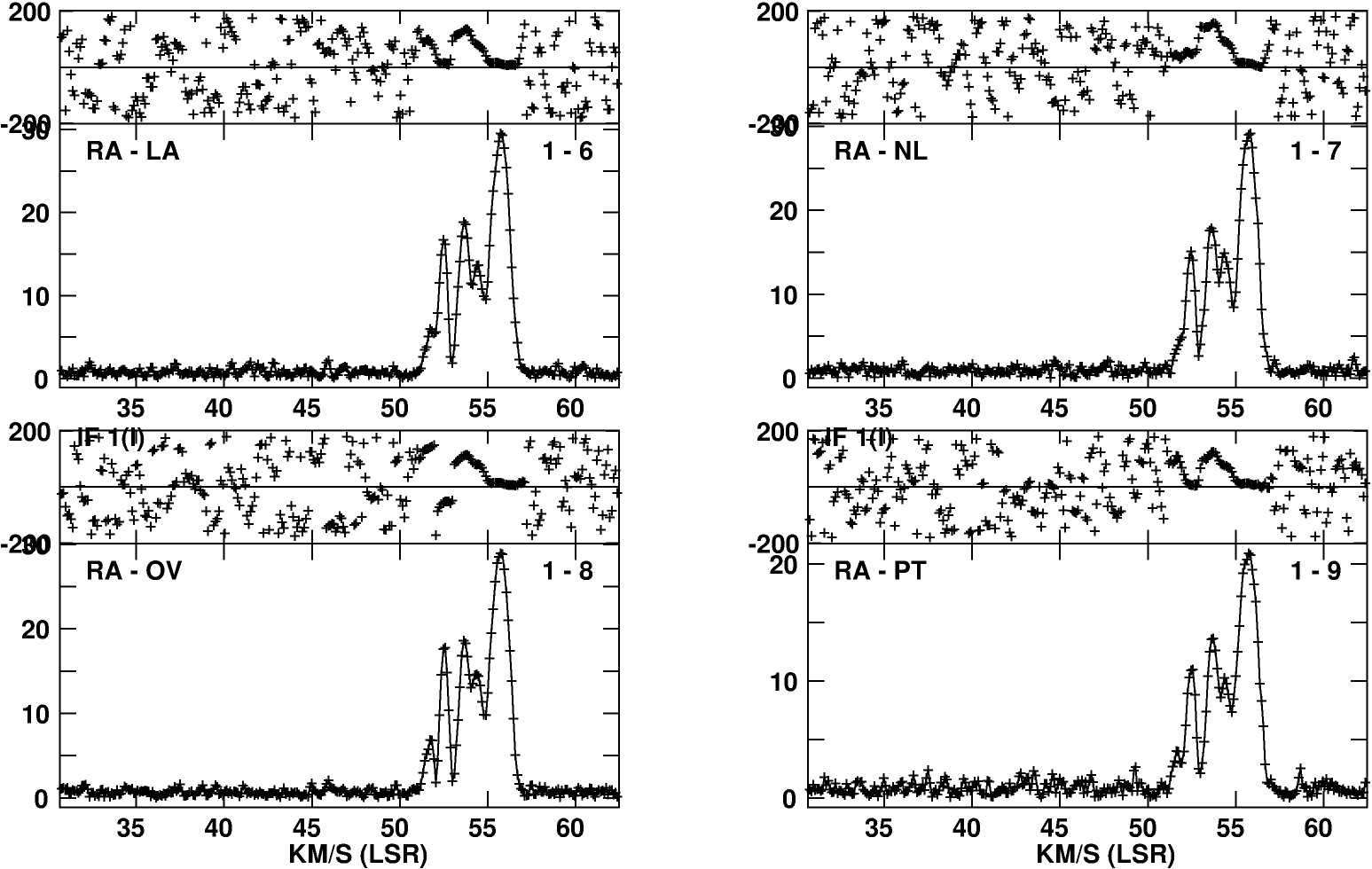}}
\end{minipage}
\caption{Stokes I vector-averaged cross-correlation spectra of G25.65+1.05 in the spectral range $+36.8\,...\,62.3$ km/s obtained on space-ground baselines between the 10-m space radio telescope and 25-m antennas of the VLBA array during Epoch I. The spectra are averaged over the first 29-min scan of the RadioAstron observation (05:31-06:00 UT). The reference spectral feature is at velocity +55.8 km/s. The amplitude is scaled in Janskies (bottom panels) and the phase is in degrees (top panels).}
\label{fig:CC_RA_full}
\end{figure*}

\begin{figure*}
\centering
\begin{minipage}[h]{0.49\textwidth}
\center{\includegraphics[scale=0.32, angle=270]{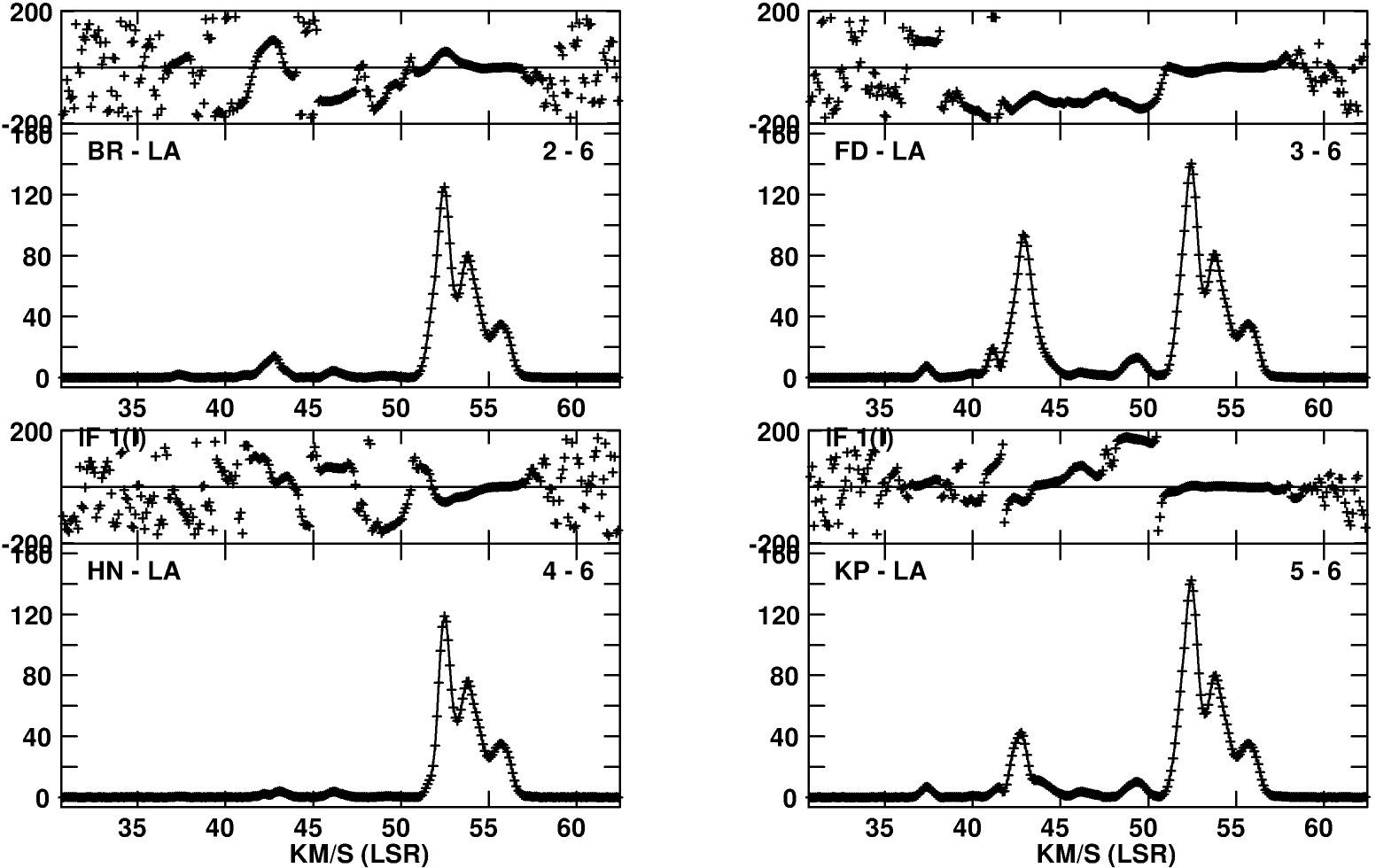}}
\end{minipage}
\hfill
\begin{minipage}[h]{0.49\textwidth}
\center{\includegraphics[scale=0.32, angle=270]{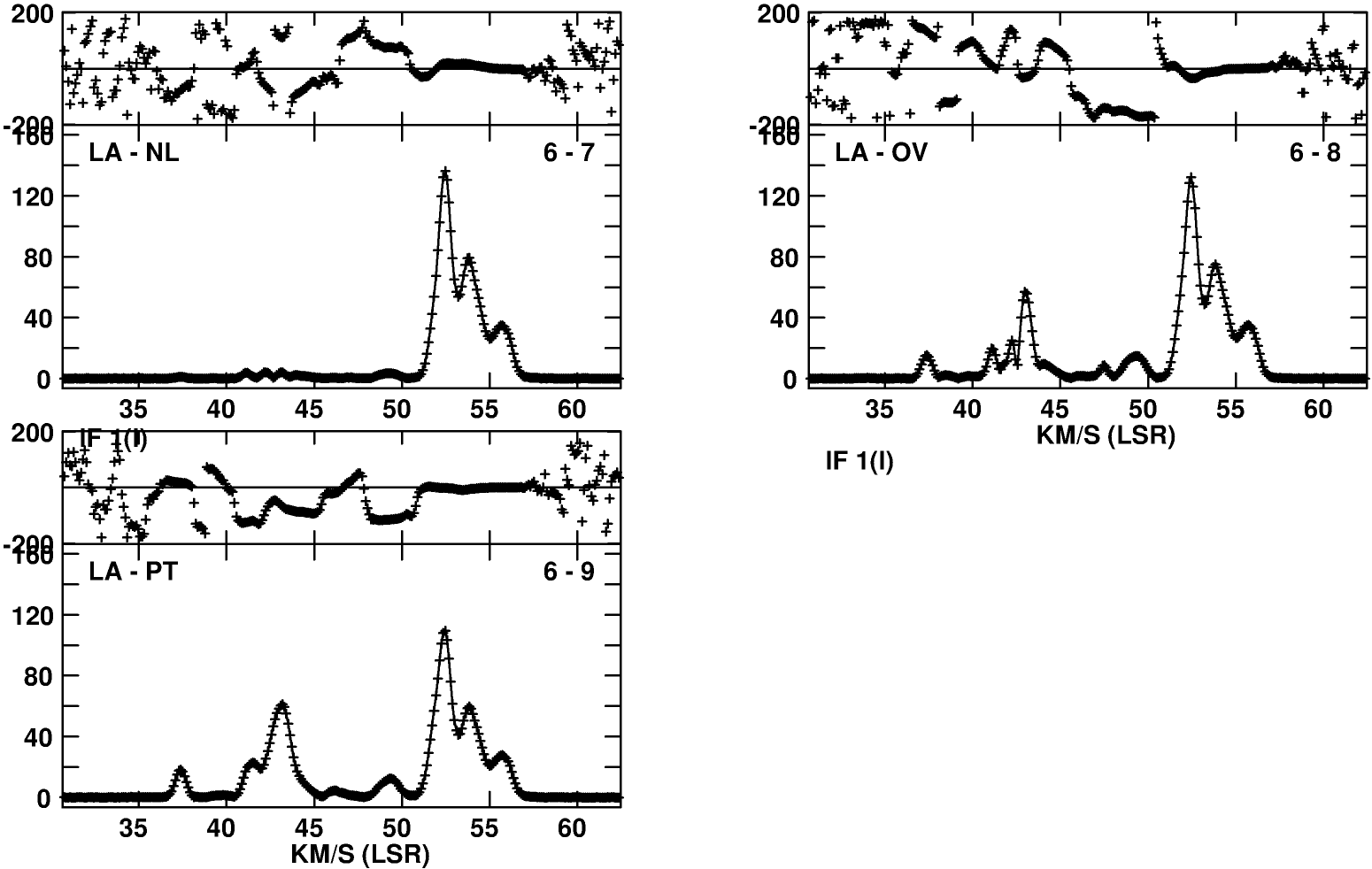}}
\end{minipage}
\caption{Stokes I vector-averaged cross-correlation spectra of G25.65+1.05 in the range $+36.8\,...\,62.3$ km/s obtained at ground baselines between the Los Alamos 25-m antenna and the remaining VLBA array antennas participated in the \texttt{rags31a} observation at Epoch~I. The spectra are averaged over the 29-min scan in time range 05:31-06:00 UT (the same as in Figure \ref{fig:CC_RA_full}). The reference spectral feature is at velocity +55.8 km/s. The amplitude is scaled in Janskies (bottom panels) and the phase is in degrees (top panels).}
\label{fig:CC_gr}
\end{figure*}

\begin{figure*}
    \centering
    \includegraphics[scale=0.4, angle=270]{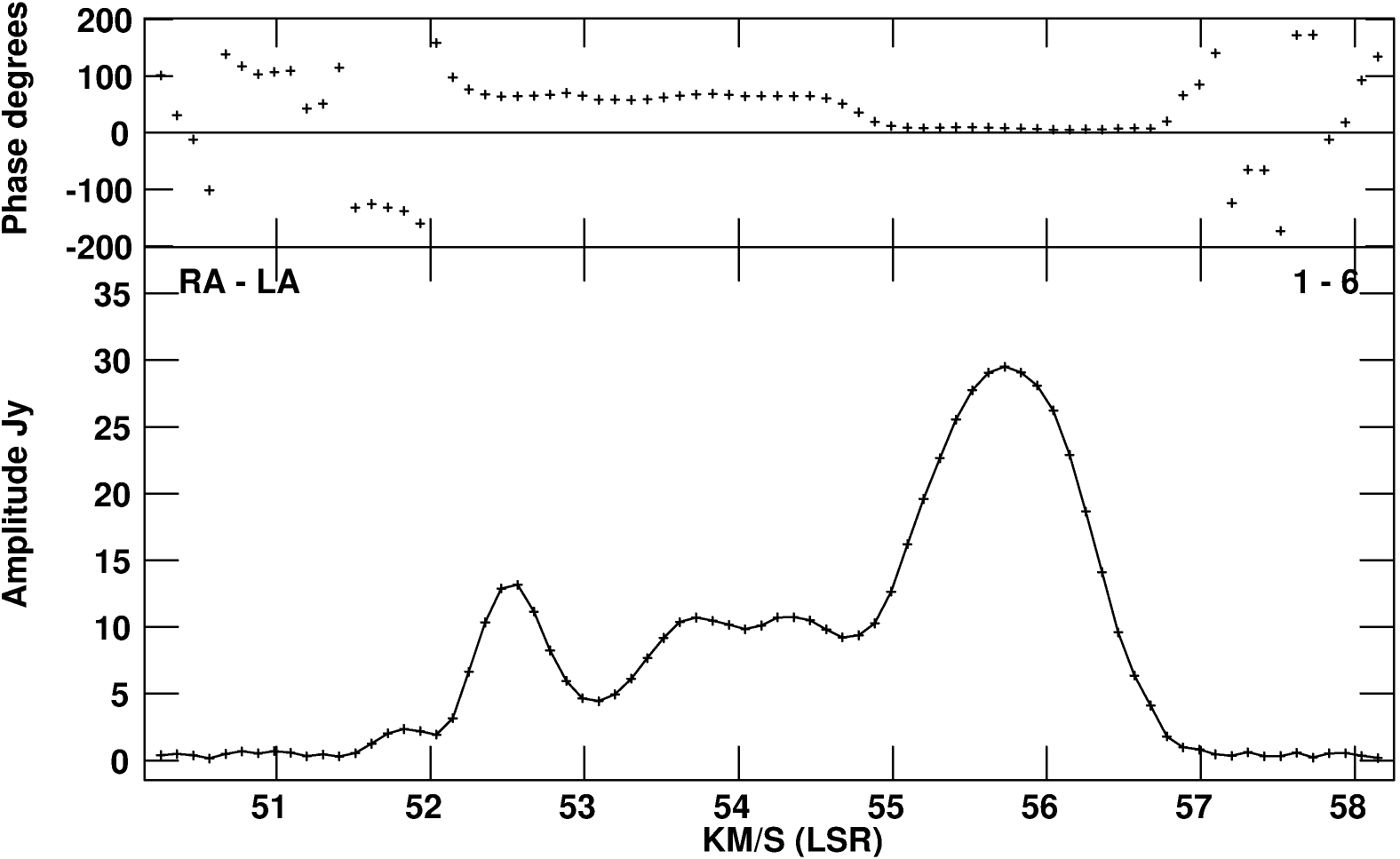}
    \caption{Stokes I vector-averaged cross-correlation spectrum in the range $+50.3\,...+58.1$ km/s obtained between the 10-m space antenna and 25-m antenna in Los Alamos during Epoch~I. The spectrum is averaged over 1.5 hours of the entire RadioAstron observation of G25.65+1.05. The reference spectral feature is the feature at +55.8 km/s. The amplitude is scaled in Janskies (bottom panels) and the phase is in degrees (top panels).}
    \label{fig:CC_RA}
\end{figure*}

\begin{figure*}
    \centering
    \includegraphics[scale=0.7]{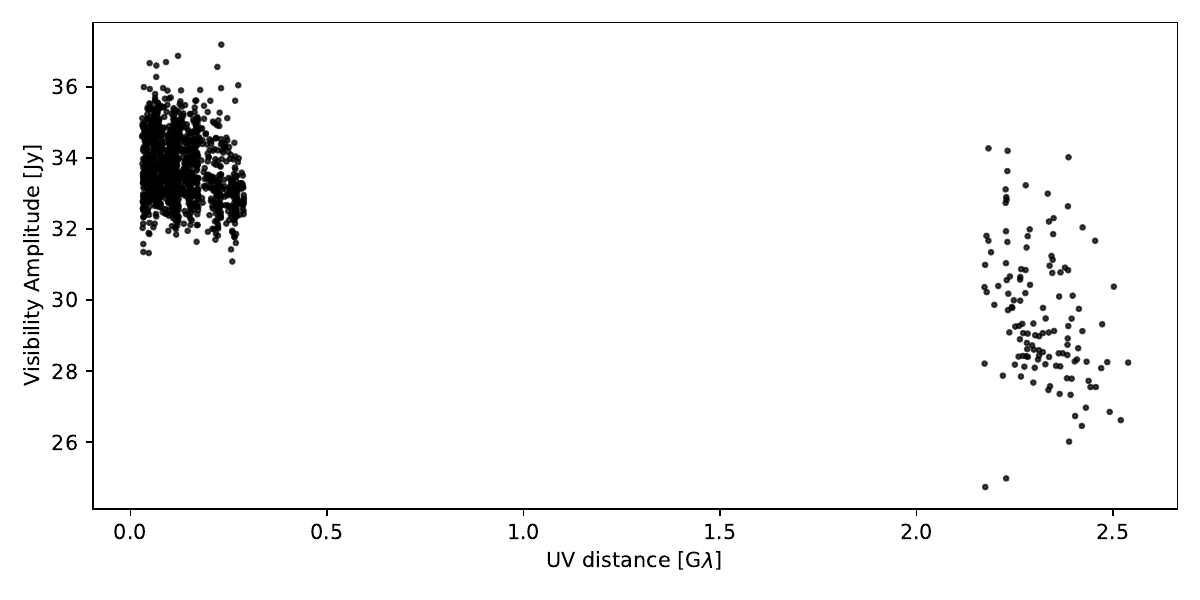}
    \caption{Correlated flux density of the most compact maser feature at +55.8 km/s depending on baseline length during Epoch~I. The data are averaged over 5 minute scans. RCP and LCP polarizations are averaged. }
    \label{fig:radplot}
\end{figure*}

H$_2$O masers in the range +51\,...\,+57~km/s were detected at both space-ground and ground-ground baselines between space antenna and all 25-m antennas of the VLBA array. Figures \ref{fig:CC_RA_full} and \ref{fig:CC_gr} shows the vector-averaged cross-correlation spectra at space-ground and ground-ground baselines in the velocity range $+36.8\,...\,62.3$ km/s comprising the existing H$_2$O maser emission in the source. The red components of the spectrum in the velocity range $+51.5\,...+57$ km/s were well detected with RadioAstron space antenna on all baselines up to 2.7 Earth diameters. Figure \ref{fig:CC_RA} represents the cross-correlation spectrum of the red features between RadioAstron and Los~Alamos averaged over the 1.5 hours of Space-VLBI observation. The maximum correlated flux density was about 30~Jy at 55.8 km/s, and this is the most compact maser feature in the source. It is appears to be almost unresolved on space baselines as seen from the plot of visibility amplitude across \texttt{UV} distance (Figure \ref{fig:radplot}). The correlated flux density restored on VLBA array is about 33~Jy, and on space-ground array it is about 30~Jy.

\begin{figure*}
\centering
\includegraphics[scale=0.5, angle=270]{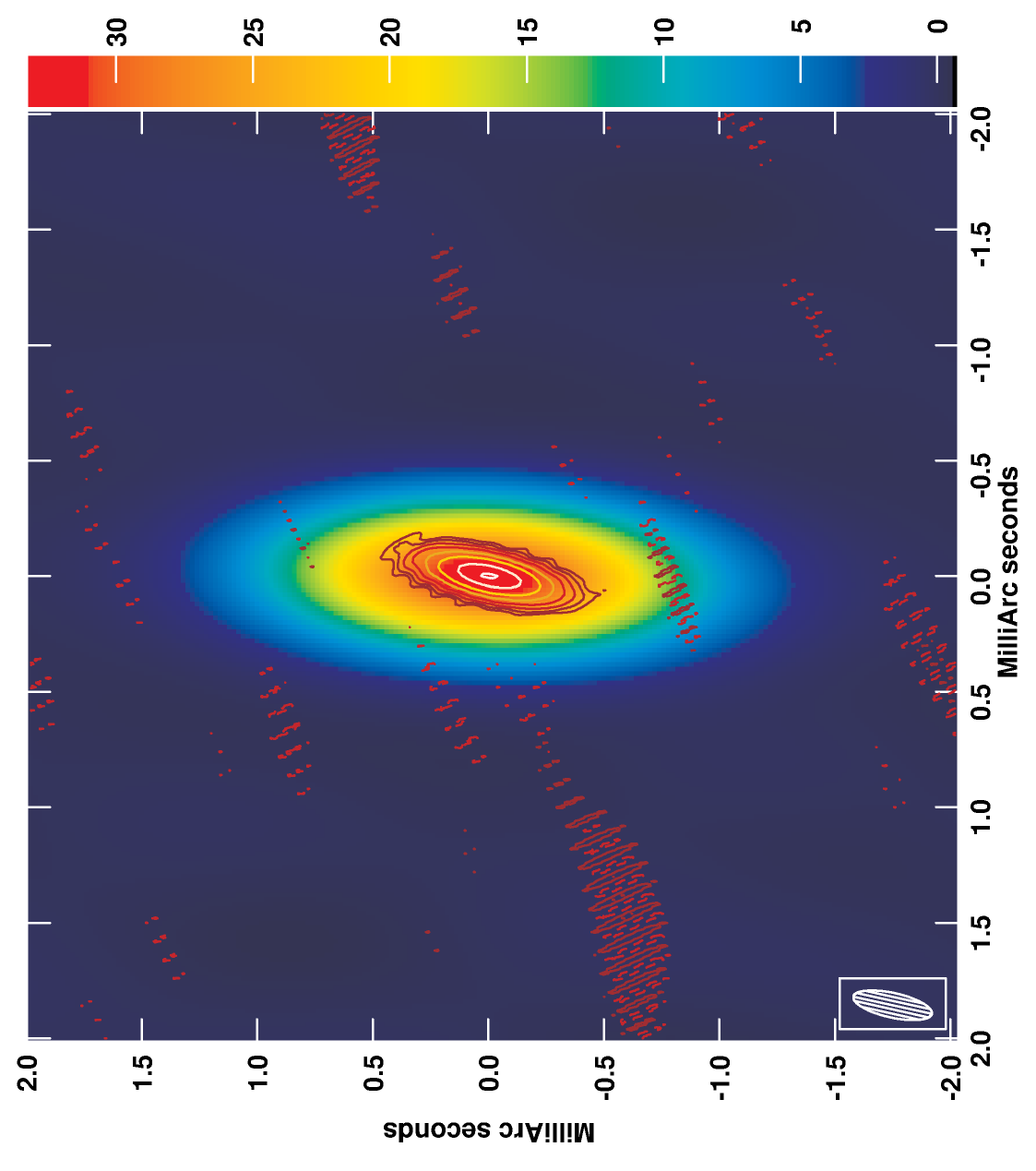}
\caption{Image of the most compact maser feature at the velocity 55.8 km/s detected on the longest RadioAstron baselines up to 2.7 Earth diameters (Epoch~I). The colored background image is obtained on the ground VLBA array during 6 hours of observation. The contours indicate the image obtained with the whole Space-VLBA array of RadioAstron during the last 1.5 hours of the observing session \texttt{rags31a}. Contour levels correspond to -1, 1, 2, 4, 8, 16, 32, 64, 96 percent of the peak flux, and the minimum and maximum contour brightness extrema are -0.632 and 33.34 Jy/beam respectively. The right vertical colored panel shows the flux scale in Jy/beam and the color scale brightness ranges from -0.65 to 33.11 Jy/beam. The Space-VLBA beam is indicated in the left bottom corner.}
\label{fig:RA_img1}
\end{figure*}

The image of the compact maser spot at +55.8 km/s, corresponding to a single spectral channel of 7.8 kHz width, is shown in the Figure~\ref{fig:RA_img1}. The colored background image is obtained on the ground VLBA array during 6 hours of observation. The contours represent the image obtained with the whole space-VLBI interferometer RadioAstron during the last 1.5 hours of observation, when space antenna was operating. Contour levels and brightness extrema are indicated in the caption of the Figure~\ref{fig:RA_img1}. 

\begin{figure*}
\centering
\includegraphics[scale=0.65]{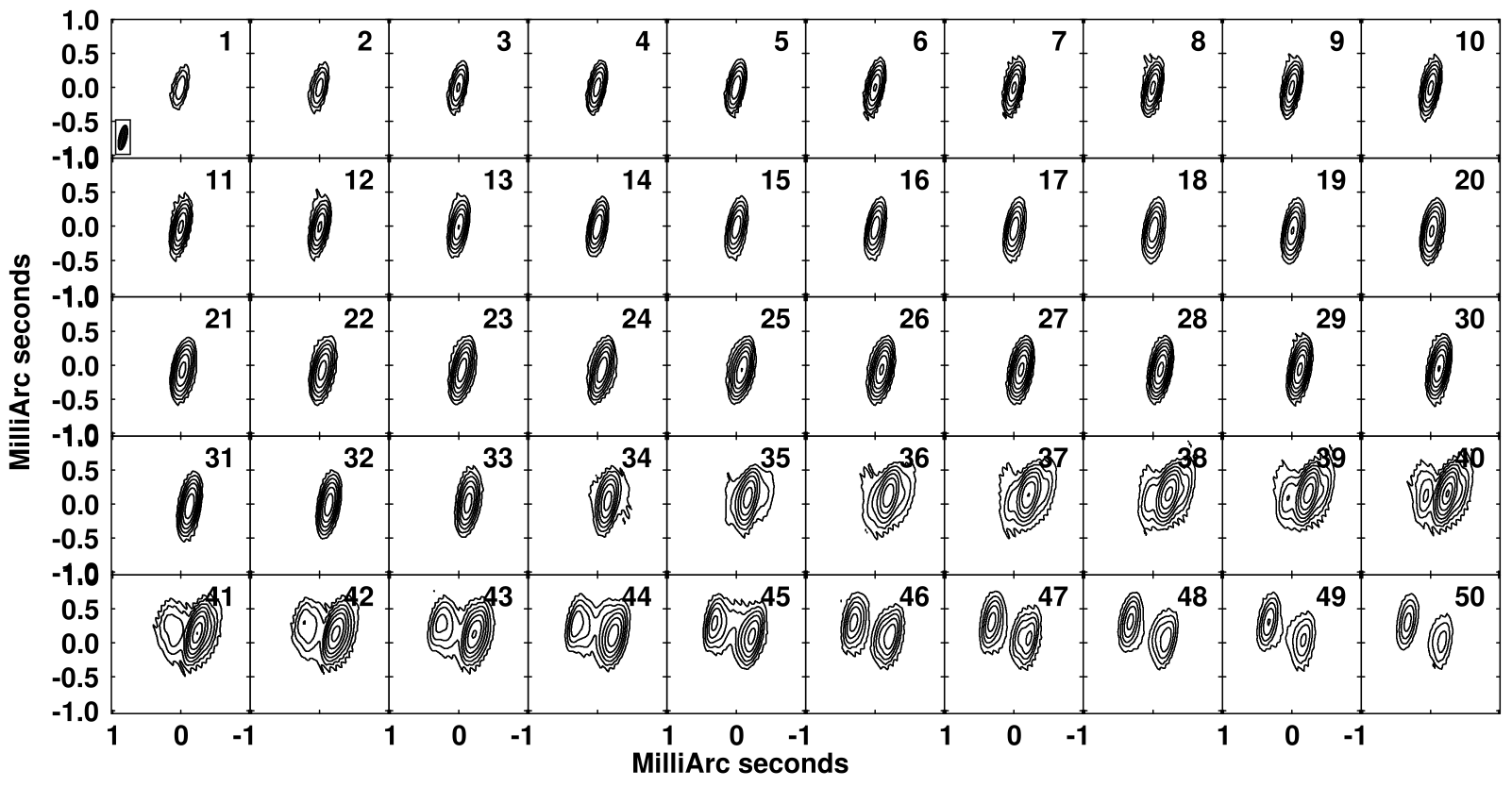}
\caption{Images of 50 spectral channels in the velocity range from +56.6 km/s (channel 1) down to +51.4~km/s (channel 50), corresponding to the red feature in the G25.65+1.05 spectrum. Contour levels are set to (-1, 1, 2, 4, 8, 16, 32, 64, 96)$\times$0.756 Jy/beam. The peak brightness is 75.6~Jy/beam in channel 40 at velocity +52.5~km/s. The images are obtained using the whole Space-VLBI array with baselines up to 2.7~Earth diameters during Epoch~I.}
\label{fig:RA_cube}
\end{figure*}

\begin{figure*}
\centering
\begin{minipage}[h]{0.49\textwidth}
\center{\includegraphics[scale=0.3, angle=270]{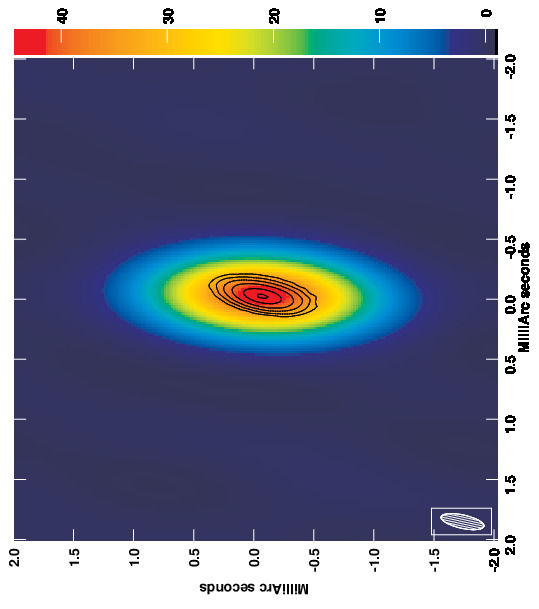}}
\end{minipage}
\hfill
\begin{minipage}[h]{0.49\textwidth}
\center{\includegraphics[scale=0.3, angle=270]{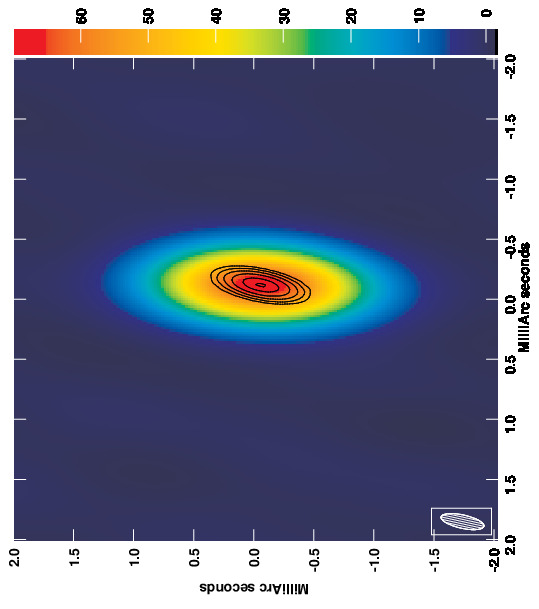}}
\end{minipage}
\vfill
\begin{minipage}[h]{0.49\textwidth}
\center{\includegraphics[scale=0.3, angle=270]{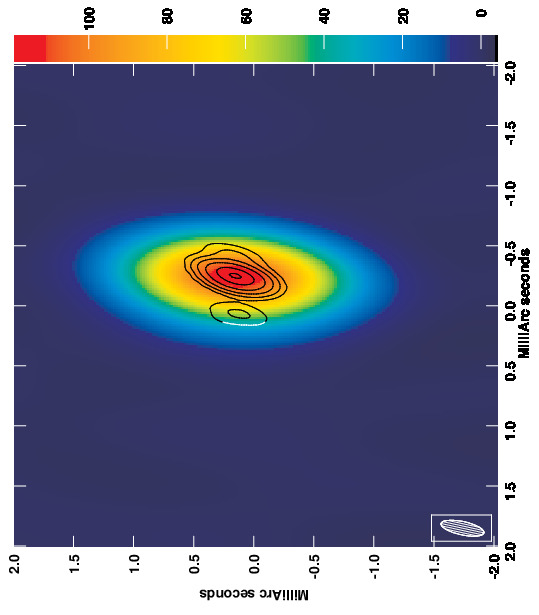}}
\end{minipage}
\hfill
\begin{minipage}[h]{0.49\textwidth}
\center{\includegraphics[scale=0.3, angle=0]{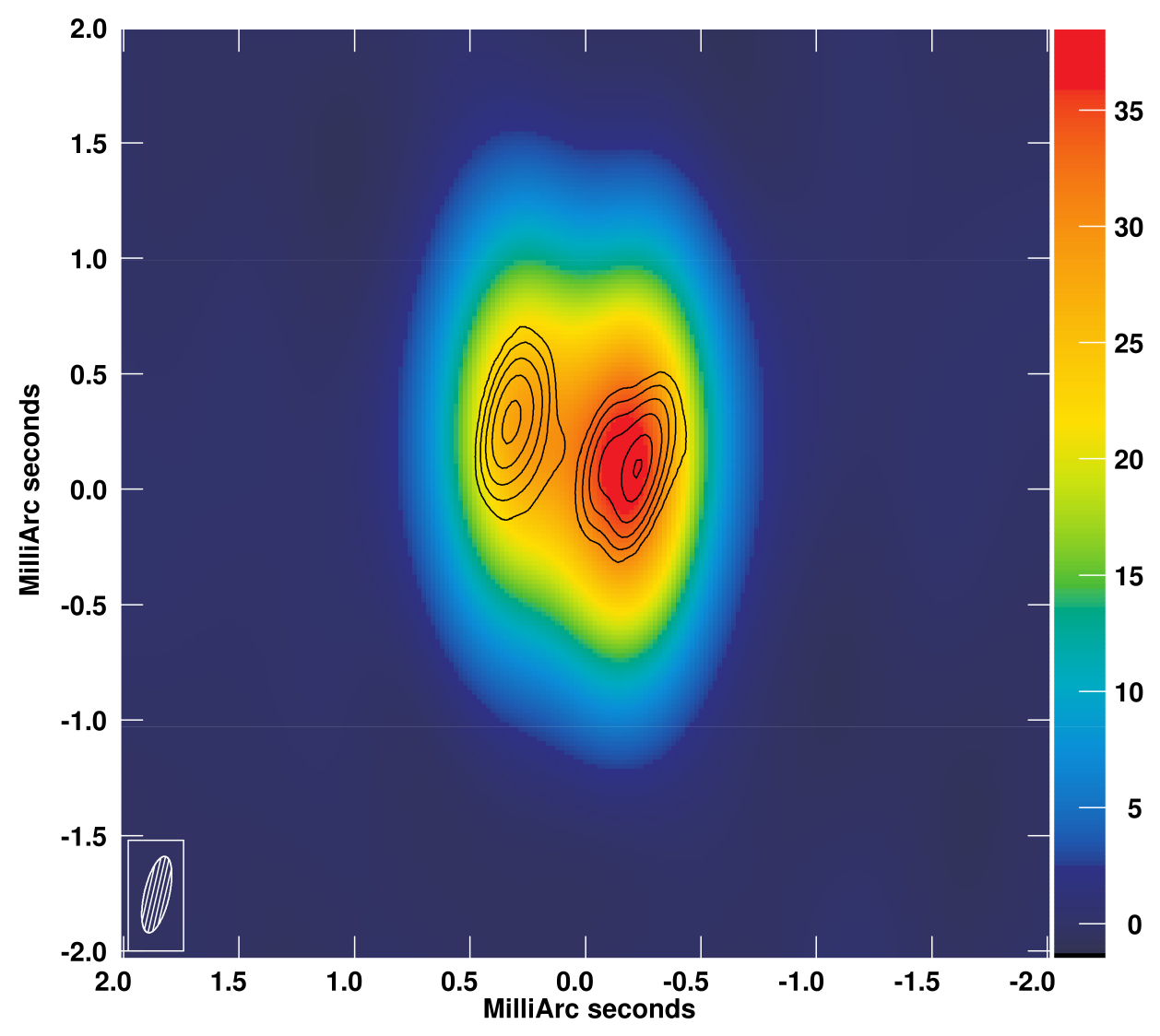}}
\end{minipage}
\caption{Images of the four maser spots corresponding to the spectral peaks at +51.8~km/s (bottom right), +52.5~km/s (bottom left), +53.6~km/s (top right), +54.5~km/s (top left), obtained with space-VLBI array RadioAstron at Epoch~I. Contour levels are set to (4, 8, 16, 32, 64, 96)~\% of the corresponding peak brightness. The colored background image is obtained on the ground VLBA array during 6 hours of observation. The brightness of the colored image in Jy/beam is indicated in color the bar on the right side of the plots.}
\label{fig:RA_peaks_img}
\end{figure*}

\begin{figure*}
\centering
\includegraphics[scale=0.4]{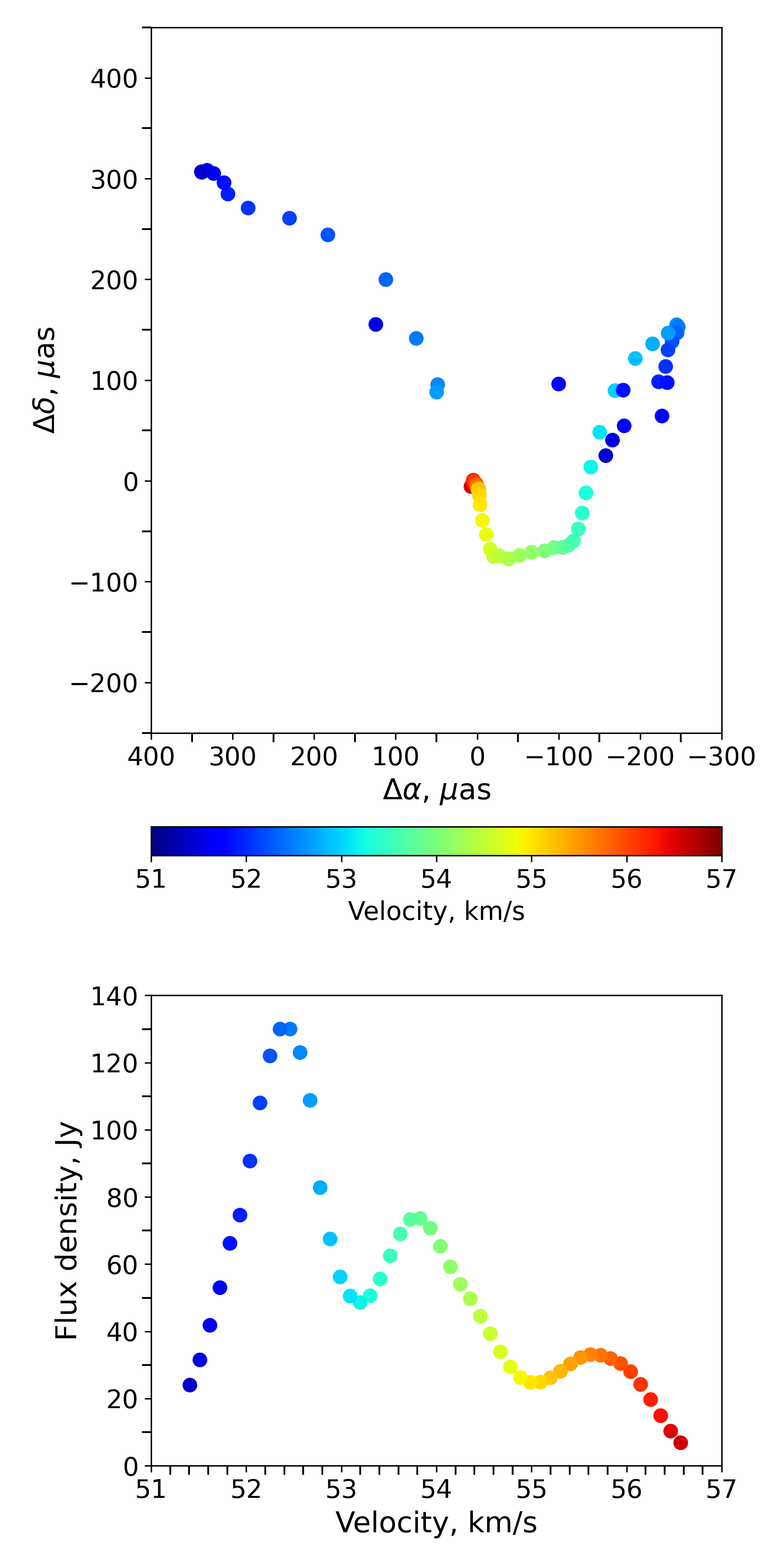}
\caption{Relative positions of H$_2$O masers spots revealed in the velocity range +51.4\,...\,+56.6~km/s (top panel) and corresponding flux density of the maser emission restored with the RadioAstron interferometer (bottom panel). The data belongs to Epoch~I.}
\label{fig:wander}
\end{figure*}

The velocity, offset position of the peak, integrated and peak flux density of each H$_2$O maser spot in the compact red spectral feature detected on space-ground baselines are listed in Table \ref{tab:jmfit_red}. Hereafter the term "maser spot" refers to a spatially distinct maser emission seen in a single spectral channel. The fitted peak positions of the spots slightly change across the velocity range +55\,...\,+57~km/s. This small drift does not exceed 20~$\mu$as across both R.A. and DEC axes, therefore all spots in this range are overlapped in the sky plane within the beam. 
The images of 50 maser spots in the whole range of red features +51.4\,...\,+56.6~km/s  detected on space baselines are presented in Figure~\ref{fig:RA_cube}. As seen from Figure~\ref{fig:CC_RA}, there are 5 distinct peaks in the cross spectrum obtained with RadioAstron at velocities 51.8, 52.5, 53.6, 54.5 and 55.8~km/s. The image of the spot corresponding to the last peak is shown on Figure~\ref{fig:RA_img1}, the images of the other four maser spots are presented in Figure~\ref{fig:RA_peaks_img}. 
Relative positions of maser spots revealed in 50 spectral channels at velocities +51.4\,...\,+56.6~km/s are presented in Figure~\ref{fig:wander}.
Possible explanation regarding the nature of these structures is discussed in Section~\ref{discussion}.

\begin{table*}
\caption{22 GHz \texorpdfstring{H$_2$O}{H2O} maser spots parameters in the most compact spectral feature detected with RadioAstron.}
\label{tab:jmfit_red}
\centering  
\begin{tabular}{cccccc}
\hline\hline
V$_{\rm LSR}$ & $\Delta\alpha$ & $\Delta\delta$ &  Peak flux & Integrated flux \\ 
(km~s$^{-1}$) & (mas) & (mas)  & (Jy/beam) & (Jy) \\
\hline
 56.57 &  -0.0076 $\pm$   0.0015 &  -0.0055 $\pm$   0.0036 &   5.30 $\pm$  0.12 &   6.83 $\pm$  0.24 \\ 
 56.46 &  -0.0054 $\pm$   0.0009 &  -0.0023 $\pm$   0.0023 &   8.45 $\pm$  0.12 &  10.30 $\pm$  0.24 \\ 
 56.36 &  -0.0045 $\pm$   0.0006 &   0.0005 $\pm$   0.0016 &  13.00 $\pm$  0.13 &  14.90 $\pm$  0.24 \\ 
 56.25 &  -0.0048 $\pm$   0.0004 &   0.0007 $\pm$   0.0011 &  17.70 $\pm$  0.12 &  19.70 $\pm$  0.23 \\ 
 56.15 &  -0.0040 $\pm$   0.0003 &  -0.0008 $\pm$   0.0009 &  21.70 $\pm$  0.12 &  24.20 $\pm$  0.23 \\ 
 56.04 &  -0.0015 $\pm$   0.0003 &  -0.0023 $\pm$   0.0008 &  25.60 $\pm$  0.14 &  28.00 $\pm$  0.25 \\ 
 55.94 &  -0.0011 $\pm$   0.0003 &  -0.0030 $\pm$   0.0008 &  28.00 $\pm$  0.14 &  30.40 $\pm$  0.26 \\ 
 55.83 &  -0.0007 $\pm$   0.0003 &  -0.0037 $\pm$   0.0007 &  29.40 $\pm$  0.14 &  31.90 $\pm$  0.25 \\ 
 55.73 &  -0.0001 $\pm$   0.0003 &  -0.0058 $\pm$   0.0008 &  30.20 $\pm$  0.14 &  32.90 $\pm$  0.26 \\ 
 55.62 &   0.0005 $\pm$   0.0003 &  -0.0077 $\pm$   0.0008 &  30.00 $\pm$  0.15 &  33.10 $\pm$  0.28 \\ 
 55.52 &   0.0009 $\pm$   0.0003 &  -0.0086 $\pm$   0.0009 &  28.90 $\pm$  0.16 &  32.20 $\pm$  0.29 \\ 
 55.41 &   0.0014 $\pm$   0.0003 &  -0.0092 $\pm$   0.0009 &  26.90 $\pm$  0.15 &  30.30 $\pm$  0.28 \\ 
 55.31 &   0.0020 $\pm$   0.0003 &  -0.0079 $\pm$   0.0009 &  24.20 $\pm$  0.14 &  28.10 $\pm$  0.26 \\ 
 55.20 &   0.0013 $\pm$   0.0004 &  -0.0084 $\pm$   0.0010 &  22.80 $\pm$  0.14 &  26.20 $\pm$  0.26 \\ 
 55.09 &   0.0027 $\pm$   0.0004 &  -0.0143 $\pm$   0.0011 &  20.20 $\pm$  0.14 &  24.90 $\pm$  0.27 \\ 
\hline                                   
\end{tabular}
\end{table*}

\begin{figure*}
\centering
\begin{minipage}[h]{0.32\textwidth}
\center{\includegraphics[scale=0.3]{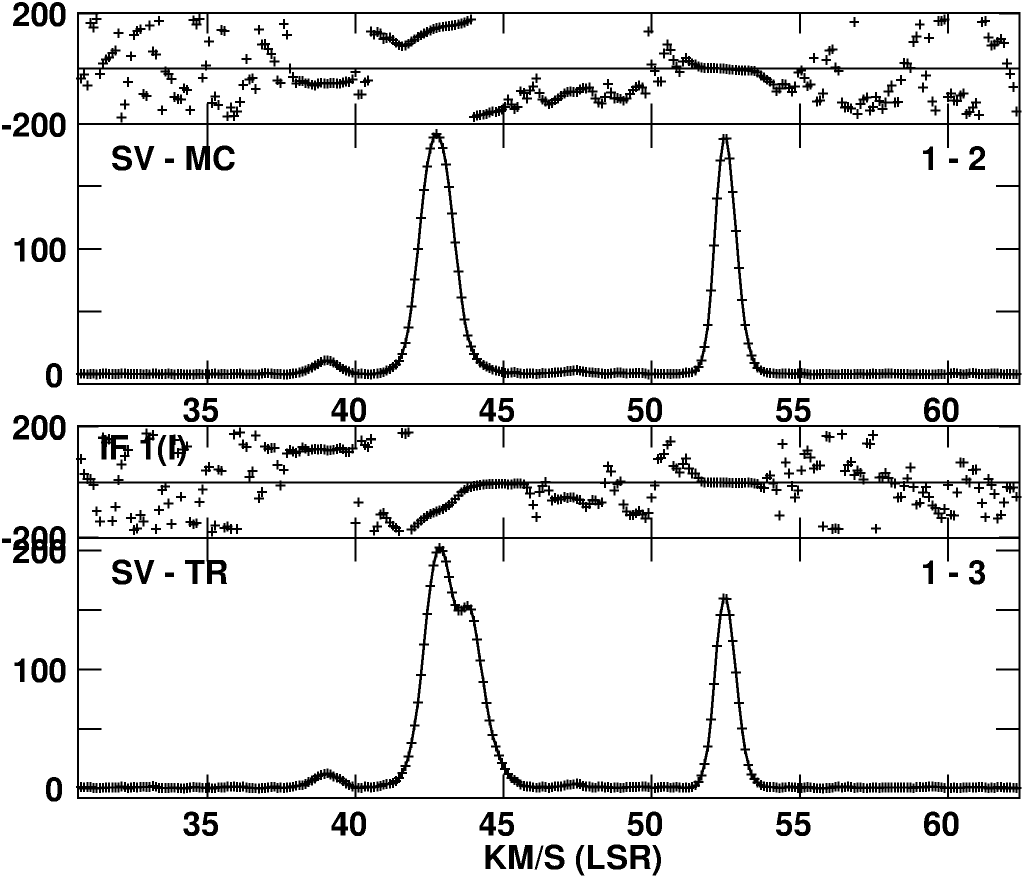}}
\end{minipage}
\hfill
\begin{minipage}[h]{0.32\textwidth}
\center{\includegraphics[scale=0.3]{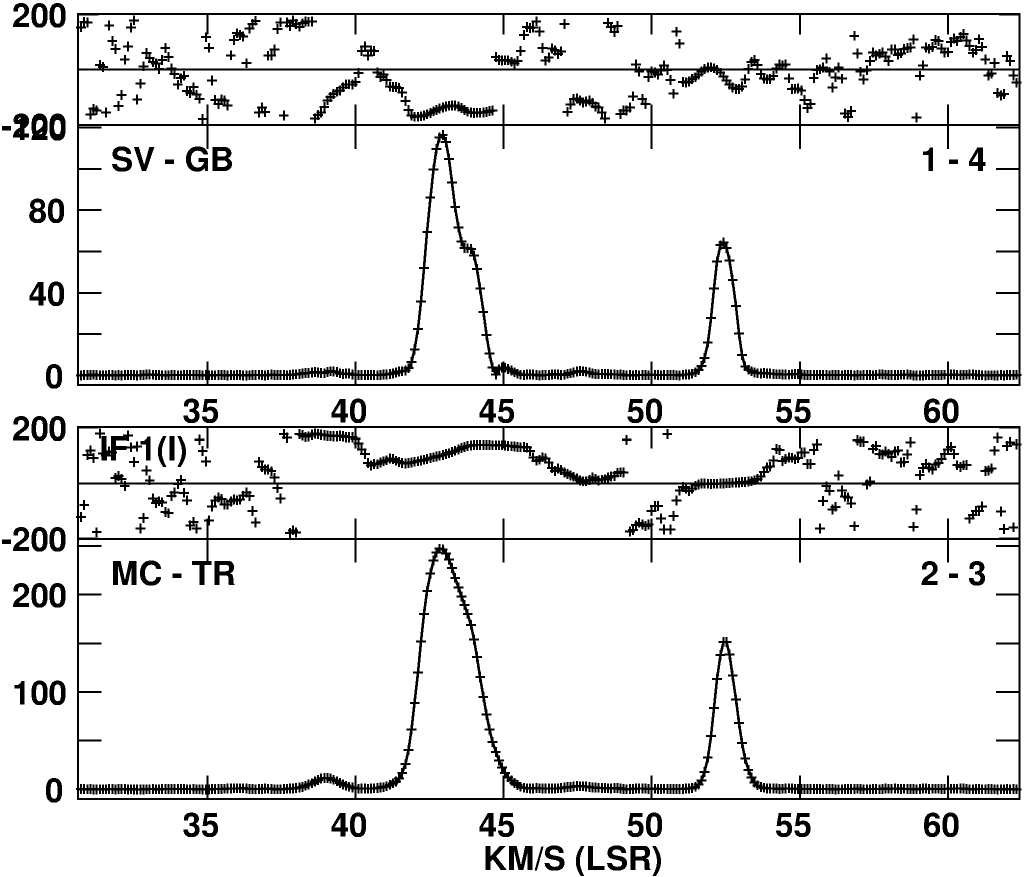}}
\end{minipage}
\hfill
\begin{minipage}[h]{0.32\textwidth}
\center{\includegraphics[scale=0.3]{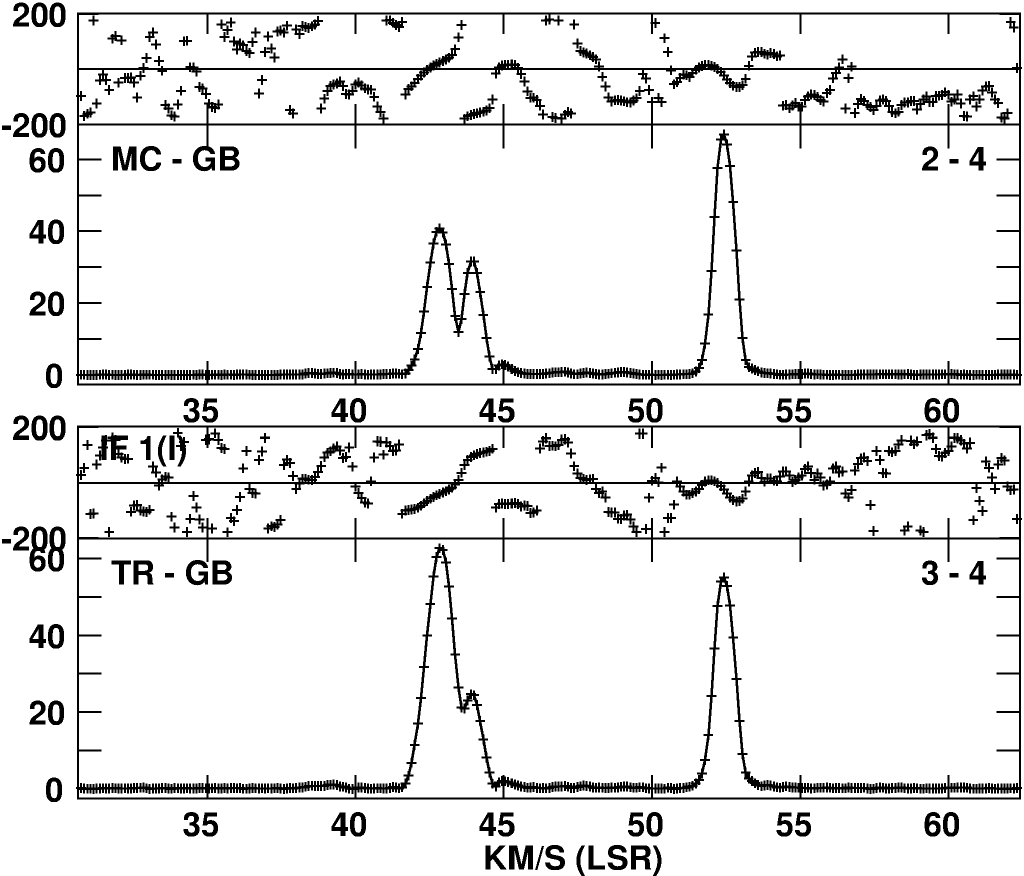}}
\end{minipage}
\caption{Stokes I vector-averaged cross-correlation spectra obtained in the \texttt{rags31b} observation during Epoch~II at ground baselines between the Green Bank, Svetloe, Torun and Medicina telescopes. The data are averaged over a 5-min scan. The reference spectral feature is the feature at 52.2 km/s. The amplitude is scaled in Jy and the phase is in degrees.}
\label{fig:CC3}
\end{figure*}

At Epoch II, 18 April 2018, red features sufficiently changed. Only a peak at +52.5~km/s survived and peaks at +53.7 and +55.8~km/s disappeared (see Figure~\ref{fig:AC}). Cross-correlation spectra in the velocity range $+36.8\,...\,62.3$ km/s obtained on ground baselines between antennas in Green Bank, Svetloe, Torun and Medicina are shown in Figure~\ref{fig:CC3}. Due to the limited \texttt{uv}-coverage in this experiment, we were not able to investigate the detailed structure of the maser in the range +51.5\,..\,+53.5~km/s. However, it appears as a single emission peak in each spectral channel rather than double peaks as it was at Epoch~I.

\subsection{Bursting \texorpdfstring{H$_2$O}{H2O} masers}

\begin{figure*}
\centering
\begin{minipage}[h]{0.49\textwidth}
\center{\includegraphics[scale=0.32, angle=270]{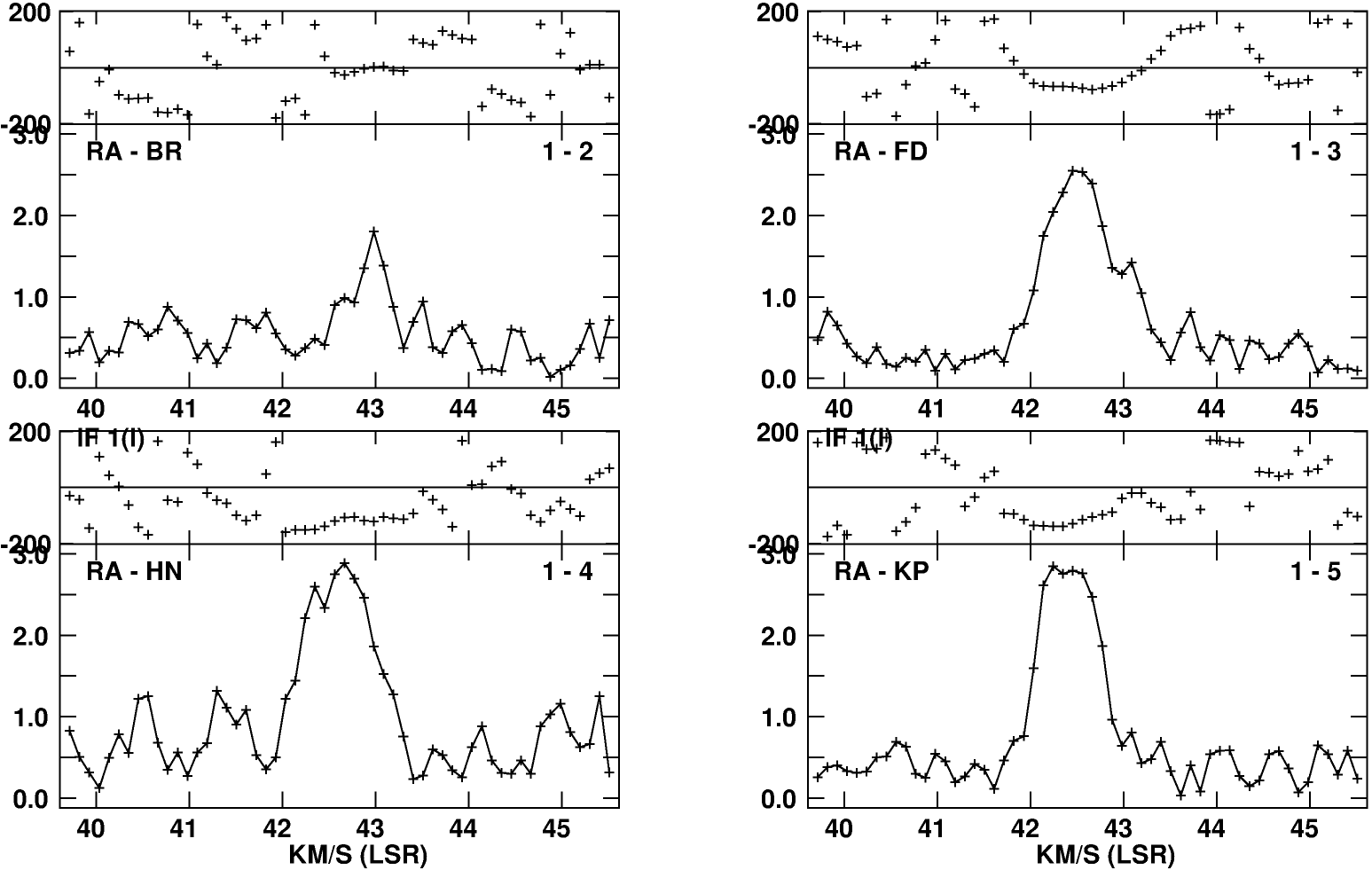}}
\end{minipage}
\hfill
\begin{minipage}[h]{0.49\textwidth}
\center{\includegraphics[scale=0.32, angle=270]{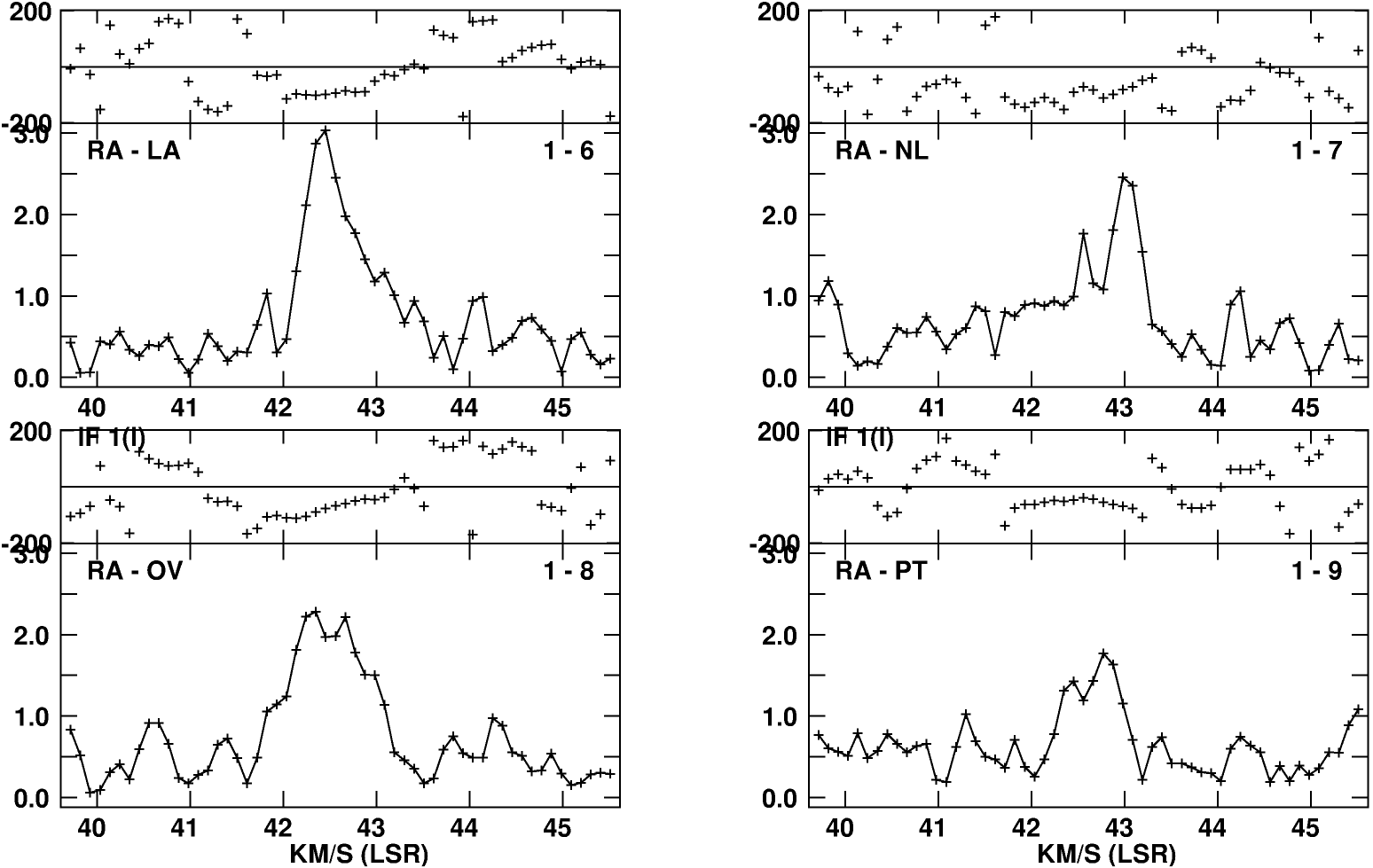}}
\end{minipage}
\caption{Stokes I vector-averaged cross-correlation spectra of G25.65+1.05 in the range of bursting feature $+39.7\,...\,45.5$ km/s obtained at space baselines between the SRT antenna and VLBA array antennas in the \texttt{rags31a} observation at Epoch~I. The spectra are averaged over the 1.5 hour of the entire RadioAstron observation. The amplitude is scaled in Jy and the phase is in degrees. The phase center is shifted to (-0.035, 0.009) arcsec from the reference feature at 55.8 km/s.}
\label{fig:CC_burst}
\end{figure*}

There is a weak correlation on space baselines with RadioAstron in the range of bursting feature. Figure~\ref{fig:CC_burst} shows the cross-correlation spectra for all space-VLBA baselines averaged over the entire 1.5 hours of the RadioAstron observation. The bursting maser feature at 42-43 km/s is offset from the reference feature at 55.8~km/s by (-0.035, 0.009) arcsec, and the spectra on Figure~\ref{fig:CC_burst} are obtained after shifting the phase center to this location. Clear detections exist on all space baselines up to 2.7~ED, however some baselines appear more noisy than others. The correlated flux density does not exceed 3~Jy of a peak at velocity 42.5~km/s. 

\begin{figure*}
\centering
\includegraphics[scale=0.6]{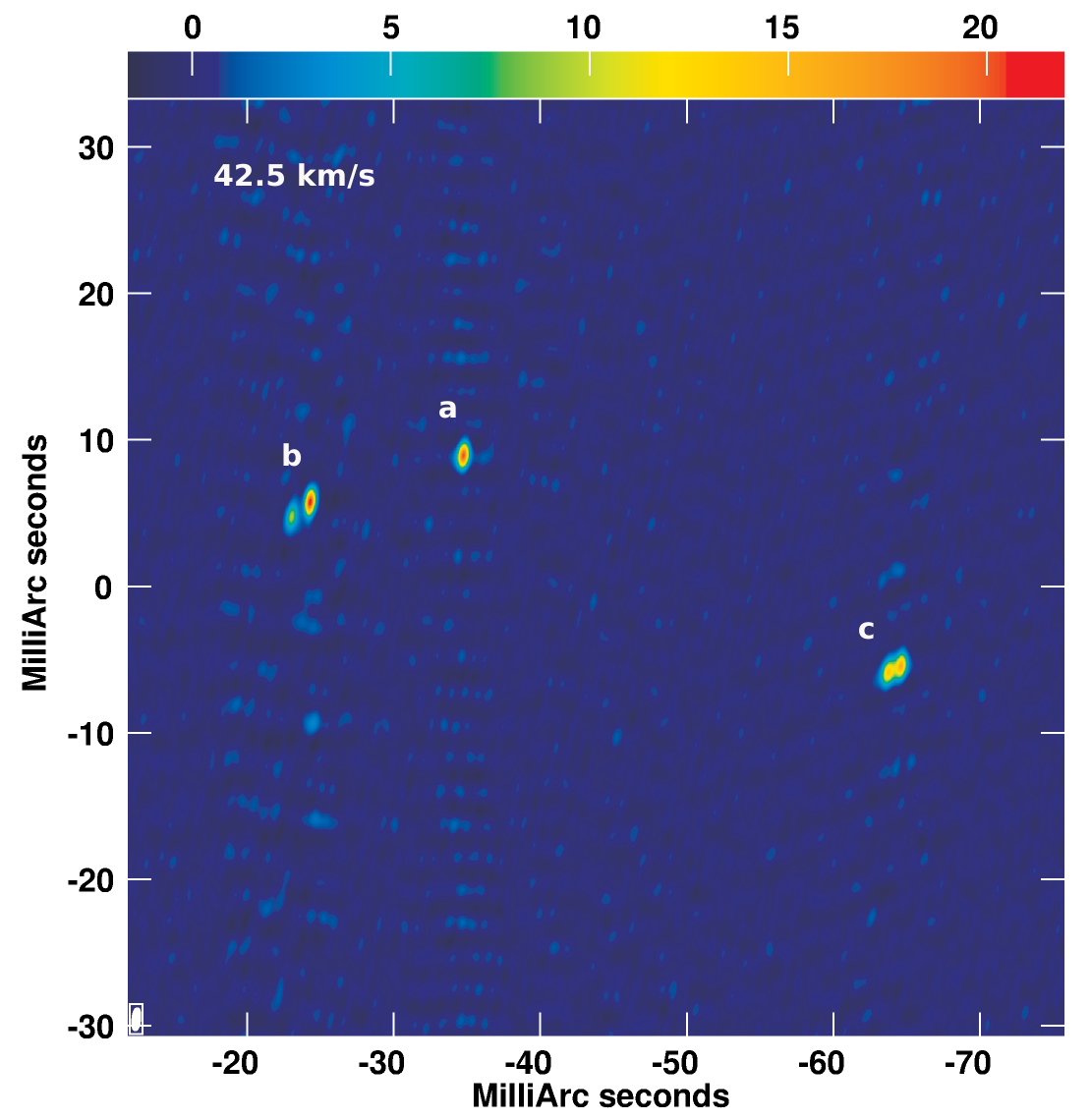}
\caption{Distribution of the maser spots in a single spectral channel of 7.8~kHz at the velocity 42.3~km/s. The map is obtained using ground VLBA array over 6 hours. The maser spot "a" at the position offset (-0.035, 0.009) arcsec is detected with RadioAstron on space baselines up to 2.7 ED at Epoch~I. The color scale brightness is indicated in the top of the figure and range from -1.64 to 21.97 Jy/beam.}
\label{fig:ch490_map}
\end{figure*}

The overall map of the emission distribution in the single spectral channel at 42.5~km/s is shown in Figure~\ref{fig:ch490_map}. There are at least five distinct maser spots in this channel at different positions: one single spot at (-0.0347, 0.009) arcsec indicated by "a" in Figure~\ref{fig:ch490_map} and two double spots indicated by "b" and "c" at (-0.024, 0.005) and (-0.064, -0.0055) arcsec, respectively. However, no correlation on space baselines was found at positions "b" and "c", only the spot "a" was detected on RadioAstron baselines. The search for detection was carried out by shifting the phase center to positions "a", "b" and "c" of maser spots revealed on ground baselines.

\begin{figure*}
\centering
\includegraphics[scale=0.4]{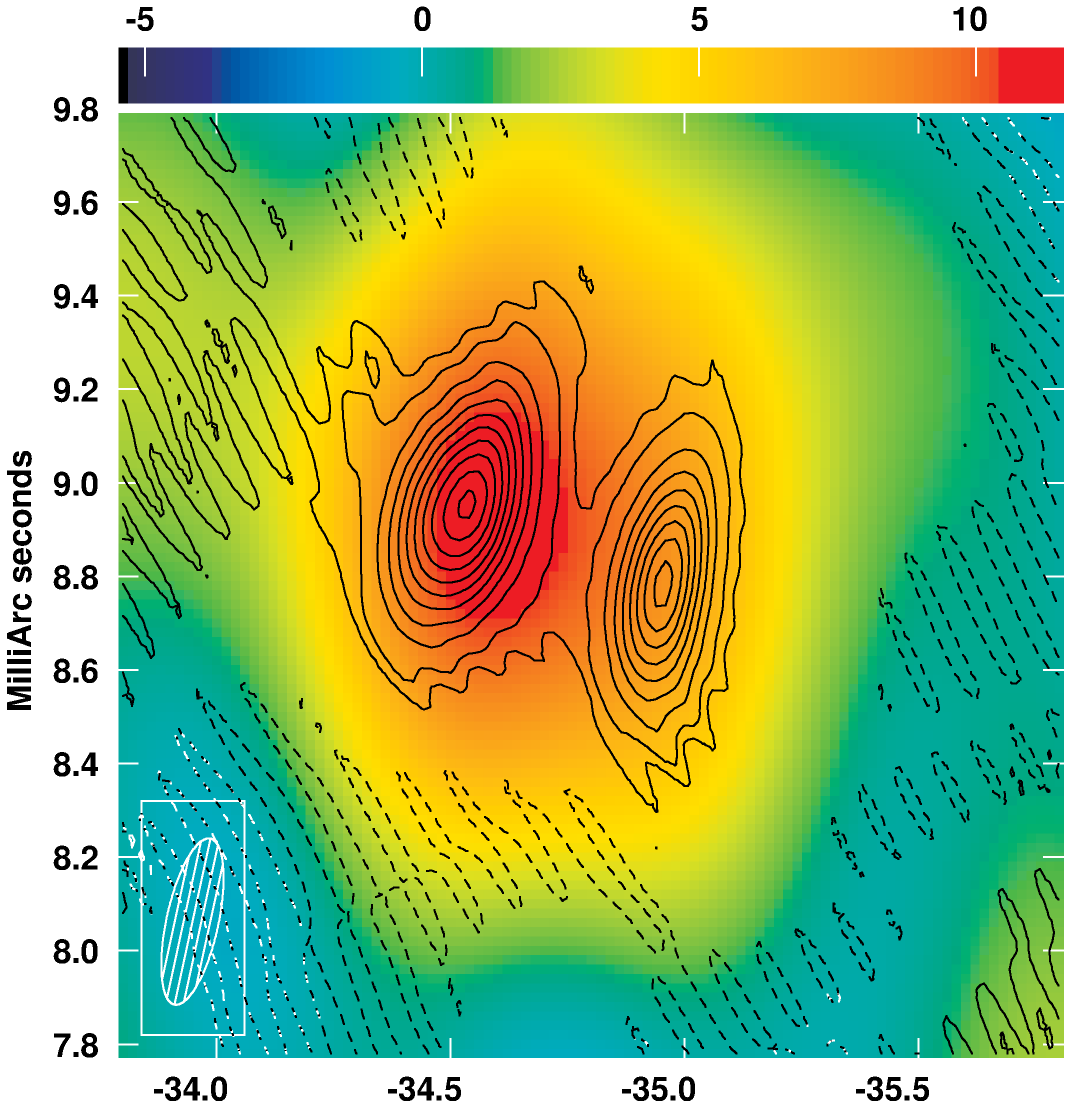}
\caption{Image of the compact maser feature at the velocity 43~km/s detected on the longest RadioAstron baselines up to 2.7 Earth diameters (Epoch~I). The colored background image is obtained on the ground VLBA array during 6 hours of observation. The contours indicate the image obtained with the whole Space-VLBA array of RadioAstron during the last 1.5 hours of the observing session \texttt{rags31a}. Contour levels correspond to (-5, 5, 10, 20, 30, 40, 50, 60, 70, 80, 90, 98)$\times$0.046 Jy/beam, and the minimum and maximum contour brightness extrema are -0.69 and 4.59 Jy/beam respectively. The colored panel shows the flux scale in Jy/beam and the color scale brightness ranges from -5.4 to 11.5 Jy/beam. The Space-VLBA beam is indicated in the left bottom corner.}
\label{fig:ch483_img_a}
\end{figure*}

Since the spot "a" was detected with RadioAstron, it is interesting to obtain an image of this spot with high angular resolution provided by Space-VLBI. At the peak velocity 42.5~km/s it appears as a single spot, however at 43~km/s the spot "a" is clearly doubled! It is important that on ground baselines this feature is not clearly resolved into double structure and only with RadioAstron baselines at higher angular resolution it is seen as two distinct spots. Figure~\ref{fig:ch483_img_a} shows the image of spot "a" at offset (-0.035, 0.009)~arcsec obtained with space interferometer (contours) along with the ground-based background image (color).

\begin{figure*}
\centering
\includegraphics[scale=0.45]{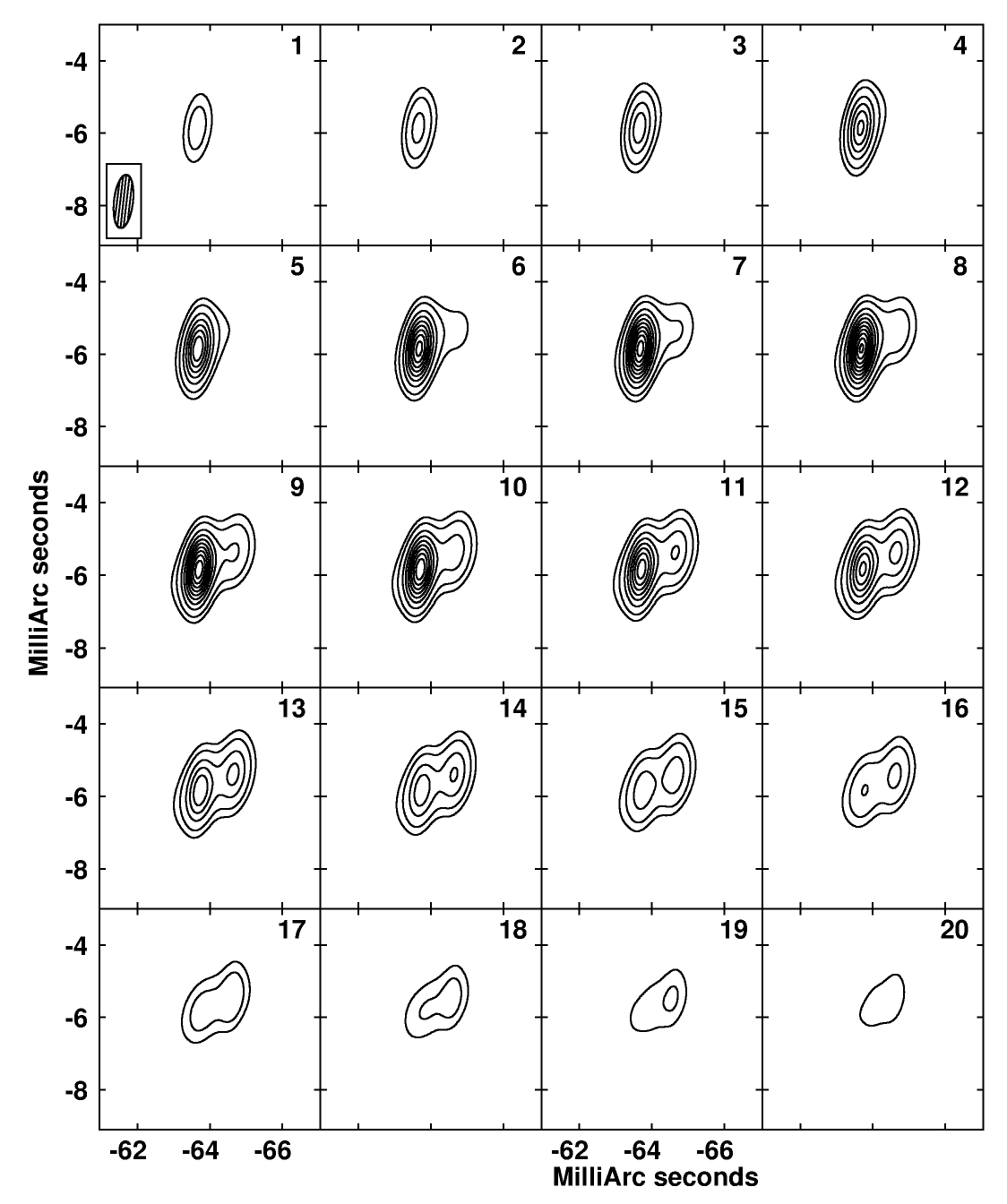}
\caption{Images of 20 spectral channels in the range 41.9\,..\,43.9~km/s, where the flaring feature in G25.65+1.05 was observed later. Contour levels are set to (5, 10, 20, 30, 40, 50, 60, 70, 80, 90, 98)$\times$0.675 Jy/beam. The peak brightness is 67.5~Jy/beam. Images are obtained at Epoch~I.} 
\label{fig:imgcube}
\end{figure*}

Due to the maser flare in G25.65+1.05 detected in this velocity range, our particular interest was focused on the maser spot "c" at the offset (-0.064, -0.005) arcsec, which is considered as a potential candidate for the source flare according to \citep{Burns2020}. It was suggested that the flare was caused by overlapping of two maser spots at this position on the sky. We were interested in obtaining images of the spots in this range and comparing them with the images obtained in \citep{Burns2020}. Image cube consisting of 50 spectral channels in velocity range 40.2\,...\,45.4~km/s was analyzed with \texttt{AIPS} task \texttt{IMAGR}. The double feature at offset (-0.064, -0.0055) arcsec is seen in 20  spectral channels in range 41.9\,..\,43.9~km/s, which is adjacent to the range 40.9\,..\,41.6~km/s of the blue wing of the bursting feature given in the Figure~4 of the paper \citep{Burns2020}. Figure \ref{fig:imgcube} shows images of these 20 spectral channels. It is clearly seen that the majority of maser spots at position (-0.064, -0.0055) have double structure. The angular offset between the two emission peaks of the double spot obtained using Gaussian fitting is about 1 mas.

At Epoch~II, the bursting feature at 42.5-43 km/s contains three maser spots at positions "a" (single spot) and "c" (double spot), and the spots at location "b" disappeared. The offsets from the reference feature, which is the same as at Epoch~I, are (-0.035; 0.009) arcsec for the single spot "a" and (-0.068; -0.0034) arcsec for the double spot "c". The angular separation between the components of the double spot "c" is about 0.36 mas, which is 2.8 times smaller than at Epoch~I. Also, the location of the double spot "c" changed by 4.3 mas between Epochs I and II over 8.3 months. Assuming that we are observing the same structure in this location that was observed during the outburst in \citep{Burns2020, Bayandina2019}, we can conclude that we are observing a shift in the crosspoint of the maser sheets. Moreover, the angular velocity of propagation of this crosspoint at the sky plane is close to the upper limit of the value 0.8$-$4.1~mas\,yr$^{-1}$ predicted in the work \citep{Burns2020}.

\subsection{Blue \texorpdfstring{H$_2$O}{H2O} masers}

The blue H$_2$O maser features at velocity +37\,..\,+39~km/s belongs to the two different regions of the source G25.65+1.05 associated with the continuum sources VLA~1 and VLA~2. This is how they differ from bursting and red features that belong only to region VLA~1. This group of blue spectral features sufficiently changed between Epochs I and II over 8.3 months. Thus, the peak at 37.4~km/s disappeared and a feature at 39.2~km/s increased almost five times from 11 to 51~Jy. Maser spots around the peak at 37.4 km/s are offset from the reference position by (0.33; -1.015) arcsec. The spots around the peak at 39.2 km/s are present both in the VLA~1 region at offset (-0.043; 0.003) arcsec and VLA~2 region at offset (0.307; -1.034) arcsec. 

The images of all these spots obtained with the VLBA array look as single emission peaks in each spectral channel. 
It is important to mention that features in this range never experienced significant flares according to single dish monitoring \citep{2018ARep...62..213L}, at least in the published observations. The fact that the other two groups of features contain double (and presumably even triple) structures suggests that the significant flares of maser emission may indeed be associated with the superposition of maser spots along the line of sight.

\subsection{Spatial distribution of maser spots}

\begin{figure*}[t]
\begin{minipage}[h]{0.49\linewidth}
\includegraphics[width=0.98\linewidth]{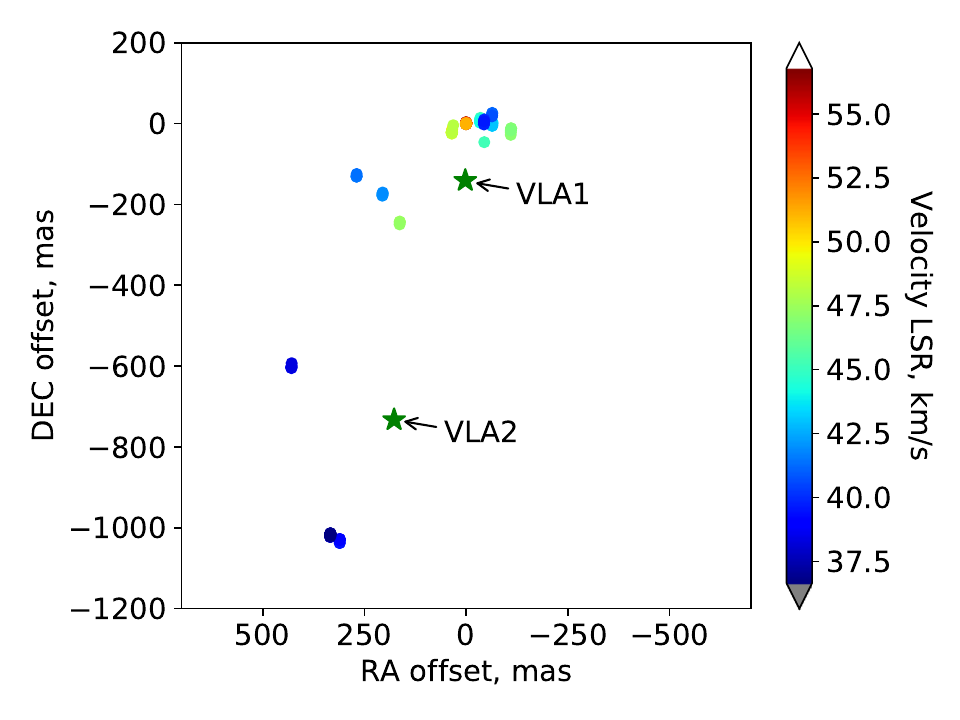}
\end{minipage}
\hfill
\begin{minipage}[h]{0.49\linewidth}
\includegraphics[width=0.98\linewidth]{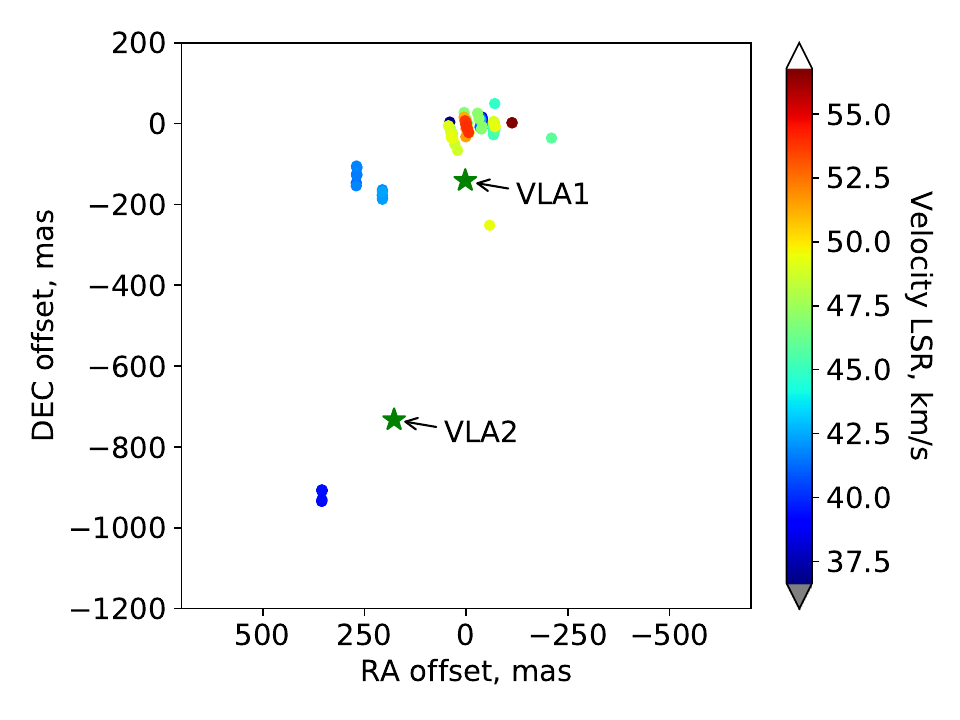}
\end{minipage}
\caption{Maps of the spatial distribution of maser emission in the RadioAstron-VLBA experiment \texttt{rags31a} at Epoch~I (left panel) and in the Epoch~II observation \texttt{rags31b} using four ground antennas in Green Bank, Svetloe, Torun and Medicina (right panel). Positions of two continuum sources VLA~1 and VLA~2 revealed in the work \citep{Bayandina2019} are indicates by stars.}
\label{fig:frmap}
\end{figure*}

The maps of the overall distribution of maser spots across all spectral channels with detectable emission corresponding to velocities in range +36\,..\,+57 km/s is presented in Figure~\ref{fig:frmap}. The \texttt{AIPS} task \texttt{FRMAP} was used to generate these maps. The search for emission in each spectral channel was conducted over a wide field, ranging from -1.75" to 1.75" in R.A. and from -2.25" to 2.25" in DEC. The reference point (0, 0) corresponds to the maser feature at the velocity +55.8 km/s. The map obtained in Epoch~I in \texttt{rags31a} experiment is presented in the left panel of Figure~\ref{fig:frmap}, and the map of Epoch~II observation \texttt{rags31b} is shown in the right panel. Positions of two continuum sources VLA~1 and VLA~2 revealed in the paper \citep{Bayandina2019} are indicated by stars in the maps. 
The maps were compared with the maser emission distribution obtained using EVN at the epoch of super flare -- see Figure~2 in the paper \citep{Burns2020}. In general, the distributions are highly similar, although some temporal changes are evident.

\section{Discussion}
\label{discussion}

We hypothesize that the H$_2$O masers observed in G25.65+1.05 originate behind a shock front. This scenario is common for H$_2$O masers in star-forming regions. By a shock front, we refer to the region where fast-moving material interacts with the ambient gas of a molecular cloud. Here, the material becomes compressed, increasing its density and temperature. Behind the shock front, the gas gradually cools down. Shock waves are ubiquitous throughout the interstellar medium and profoundly influence the star formation process. They arise, for instance, through interactions between outflows from young forming stars and their surroundings, as well as via stellar winds, supernova explosions, and collisions among molecular clouds. Shocks can occur for various reasons and exhibit diverse characteristics (such as J-shocks and C-shocks). Importantly, shocks — and consequently, their fronts — are nearly always present in star-forming regions. Observationally, these shocks can manifest as chains of H$_2$O masers. Evidence shows that H$_2$O masers frequently emerge directly behind shock fronts, where conditions favor their amplification (as supported by theoretical studies by \citet{1989ApJ...346..983E}; \citet{2013ApJ...773...70H}; and \citet{2022AstL...48..345N}. In the case of G25.65+1.05, this is highly plausible, as distinctive maser chains forming characteristic arcs are plainly discernible in our observations. These maser chains are also prominently visible in data presented by Burns (2020) and Bayandina (2019).

Based on our interferometric observations, we calculated the brightness temperatures of multiple spectral components, which are presented in columns 2 and 4 in the Table~\ref{tab:tau}. The evolution of G25's spectrum across two observing sessions, \texttt{rags31a} and \texttt{rags31b}, is illustrated in Figure \ref{fig:AC}. Notably, the spectrum underwent substantial changes over eight months: certain spectral features vanished entirely, while others shifted markedly in shape and intensity. Drawing upon theoretical models (e.g., \citet{2022AstL...48..345N}), we adopted the probable gas temperature within shock regions where H$_2$O masers develop. Given that brightness temperature can be determined by kinetic gas temperature $T{_\text{kin}}$ and optical depth $\tau$ as $T_B = T{_\text{kin}}\times e^{|\tau|}$, we calculated the optical depth in the center of the maser lines. The results are presented in the Table \ref{tab:tau} in columns 3 and 5, which correspond to Epochs~I and II. 

Focusing on the brightest feature at 43 km/s, the optical depths measured were $\tau=19.3$ in session \texttt{rags31a} and $\tau=23.5$ in session \texttt{rags31b}. Consequently, the optical depth increased by a factor of 1.2 over eight months. How might one interpret this change? We should consider that the maser's optical depth scales with the gain path length along the line of sight but diminishes with the thickness of the shock wave (assuming a uniform gain per unit volume). One potential explanation involves a 1.2-fold elongation of the path length, possibly resulting from a slight reorientation of the shock front relative to the observer. As the optical depth correlates positively with path length, an edge-on perspective would yield higher values, leading to more intense maser emissions. This model accounts for the rapid variability and high luminosities of H$_2$O masers. Alternatively, the shock front may have contracted by a factor of 1.2 between observation periods, enhancing the particle density and thereby boosting the gain. Lastly, a hybrid scenario combining both factors cannot be ruled out.

\begin{table*}
\centering  
\begin{threeparttable}
\caption{Maser line depth for different peaks in G25.65+1.05}
\label{tab:tau}
\begin{tabular}{ccccc}
\hline\hline   
V$_{\rm LSR}$ & T$_{\rm B1}$ &  |$\tau_1$| & T$_{\rm B2}$ & |$\tau_2$| \\ 
(km~s$^{-1}$) & (K) &   & (K) &  \\
\hline
 37.36 & 1.15e+12 & 20.86 &  1.44e+10 & 16.48 \\
 39.15 & 2.30e+10 & 16.95 &  7.61e+11 & 20.45 \\
 41.57 & 9.96e+10 & 18.42 &  1.28e+12 & 20.97 \\
 42.31 & 6.72e+10 & 18.02 &  9.54e+12 & 22.98 \\
 42.84 & 1.93e+11 & 19.08 &  1.61e+13 & 23.50 \\
 43.05 & 2.33e+11 & 19.27 &  1.53e+13 & 23.45 \\
 52.42 & 1.75e+12 & 21.28 &  8.19e+12 & 22.83 \\
 53.68 & 5.79e+11 & 20.18 &  1.72e+11 & 18.96 \\
 55.58 & 5.16e+12 & 22.36 &  2.93e+10 & 17.19 \\
\hline                                   
\end{tabular}
\begin{tablenotes}\footnotesize
\item[] V$_{LSR}$ is the line peak velocity, T$_{B1}$ and T$_{B2}$ -- line brightness temperatures in rags31a and rags31b sessions respectively, $\tau_1$ and $\tau_2$ are the line depth in \texttt{rags31a} and \texttt{rags31b} sessions.
\end{tablenotes}
\end{threeparttable}
\end{table*}

The figure \ref{fig:imgcube} displays images across multiple spectral channels for the flaring feature, with velocities from 41.9 to 43.9 km/s.
In paper \citep{Burns2020}, double structure of the masers was observed at velocities 40.9\,...\,41.6~km/s. Therefore, the velocity ranges of double feature in two different epochs appear to form a continuous sequence. Evidently, within the almost two months between our pre-flare observations and session at the flare epoch, either one or both features experienced a shift in velocity. However, most importantly, at a velocity of 41.6$-$41.7 km/s, the images from both epochs resemble a single feature, considering the differing beams. This observation potentially supports the hypothesis suggesting an overlap of two features, leading to enhanced radiation and the flare.

The nature of the most compact "red" feature might be different from the main spectral component at 43 km/s. The maser spots within this feature form an ordered curved chain as shown in Figure~\ref{fig:wander}. Spots at velocities 52.5\,..56.7 km/s are single-peaked and spots at 51.3\,..\,52.5 km/s are double-peaked and form two separate chains of masers. 

The spots responsible for the most compact red feature at 55.8 km/s forms a small arc structure (red end of the chain), which is displaying an apparent angular scale of approximately 60~$\mu$as. The linear scale of this projected structure is 0.15 a.u. at the distance 2.5 kpc adopted from \citep{2007PASJ...59.1185S}. We propose that this configuration could represent a rotating turbulent vortice, estimated to have a linear size of roughly 0.18 a.u., considering the statistically averaged possible inclination angle of $\sim$55$^\circ$ between the structure and the line of sight. This translates to a scale equivalent to 18 solar diameters. Such turbulent cells can arise in the interstellar medium and they have very compact sizes. For example, in \citep{2018ApJ...856...60S} such vortice was discovered in star-forming region Cepheus~A and detected on space-ground baselines of RadioAstron interferometer up to 3 Earth diameters. Another explanation of the observed structure that can be hypothesized is that these compact masers trace a disk around hidden low-mass protostellar object, which had not been revealed as continuum source yet.

\section{Conclusions}

\begin{itemize}

\item[$\star$] A super-compact H$_2$O masers in G25.65+1.05 were detected in the velocity range from 51.4 to 56.6 km/s using Space-VLBI array RadioAstron. The detections were obtained on all space-ground baselines up to 2.5~M$\lambda$. The most compact H$_2$O maser is revealed at 55.8~km/s and unresolved on baselines up to 2.5 M$\lambda$. The brightness temperature of the most compact maser is 5.2$\times10^{12}$~K. The angular size of the compact maser structure is 60 $\mu$as and a corresponding linear size is 0.18 a.u. at a distance of 2.5 kpc, which is comparable to the size of the Sun. This compact structure underwent significant changes in the spectrum over the 8 months between the two observation sessions, but its spatial location remained constant.

\item[$\star$] The maser feature at 43 km/s responsible for the super-flares in 2016-2017 had fluxes of 200 and 400 Jy at the two observing sessions, respectively, i.e. the maser was in a quiet state during both epochs. This feature contains maser emission in several distinct spatial components across the sky, one of which is detected on Space-VLBI baselines with correlated flux density of $\sim$3~Jy and has a double structure.

\item[$\star$] A flare model is proposed based on the analysis of channel-by-channel images of maser spots corresponding to the spectral range of 42\,...\,44 km/s. We confirm that the flaring feature had a double structure in the pre-flare epoch, but the feature was present on slightly different velocities comparing to the flare epoch, with a difference of about 1~km/s. This is in agreement with idea of moving sheets, which finally overlapped in the line of sight and caused a flare. The angular shift of the double flaring maser spot between the two epochs is 4.3 mas over 8.3 months.  

\item[$\star$] The maps of maser emission spatial distribution across the sky with an interval of 8 months were obtained. The overall structure of the H$_2$O masers in the region on a large scale has not undergone significant changes. However, there are some differences for a group of maser spots at an angular distance of $\sim$1 arcsec from the phase center. A group of masers around velocity 37.4 km/s disappeared both in the spectrum and in the map at Epoch~II, while the group at 39~km/s increased in flux at Epoch II and present at both epochs. 

\item[$\star$] The brightness temperatures of maser features and the optical depth in the center of maser lines were obtained. A comparison of these parameters between two observational epochs revealed that the optical depth of the brightest bursting feature increased by factor 1.2, which is interpreted as (1) an increase in gain due to a change in the geometric path on the line of sight, or (2) enhancing the particle density and boosting the gain, or both scenarios might contribute.

\end{itemize}

\section*{Acknowledgments}

The RadioAstron project is led by the Astro-Space Center of the P.N.Lebedev Physical Institute of the Russian Academy of Sciences and the Lavochkin Scientific and Production Association under a contract with the State Space Corporation ROSCOSMOS, in collaboration with partner organizations in Russia and other countries.

The Long Baseline Observatory is a facility of the National Science Foundation operated under cooperative agreement by Associated Universities, Inc.

\section*{Data Availability}

The data underlying this article are available by request to the authors.

\bibliographystyle{jasr-model5-names}
\biboptions{authoryear}
\bibliography{refs}

\begin{thebibliography}{33}
\expandafter\ifx\csname natexlab\endcsname\relax\def\natexlab#1{#1}\fi
\ifx\xfnm\relax \def\xfnm[#1]{\unskip,\space#1}\fi

\bibitem[{{Ashimbaeva} et~al.(2020){Ashimbaeva}, {Krasnov}, {Lekht},
  {Pashchenko}, {Rudnitskii} \& {Tolmachev}}]{2020ARep...64...15A}
\bibinfo{author}{{Ashimbaeva}, N.~T.}, \bibinfo{author}{{Krasnov}, V.~V.},
  \bibinfo{author}{{Lekht}, E.~E.} et~al. (\bibinfo{year}{2020}).
\newblock \bibinfo{title}{{Structure and Evolution of Powerful H$_{2}$O Maser
  Flares in the Protostellar Object IRAS 18316-0602 (G25.65+1.05)}}.
\newblock {\it \bibinfo{journal}{Astronomy Reports}\/},  {\it
  \bibinfo{volume}{64}\/}\bibinfo{issue}{(1)}, \bibinfo{pages}{15--22}.
  \DOIprefix\doi{10.1134/S1063772920010011}.

\bibitem[{{Ashimbaeva} et~al.(2017){Ashimbaeva}, {Platonov}, {Rudnitskij} \&
  {Tolmachev}}]{2017ATel11042....1A}
\bibinfo{author}{{Ashimbaeva}, N.~T.}, \bibinfo{author}{{Platonov}, M.~A.},
  \bibinfo{author}{{Rudnitskij}, G.~M.} et~al. (\bibinfo{year}{2017}).
\newblock \bibinfo{title}{{The H2O Maser G25.65+1.05 Flared Again}}.
\newblock {\it \bibinfo{journal}{The Astronomer's Telegram}\/},  {\it
  \bibinfo{volume}{11042}\/}, \bibinfo{pages}{1}.

\bibitem[{{Bayandina} et~al.(2023){Bayandina}, {Burns}, {Kurtz}, {Moscadelli},
  {Sobolev}, {Stecklum} \& {Val'tts}}]{Bayandina2023}
\bibinfo{author}{{Bayandina}, O.~S.}, \bibinfo{author}{{Burns}, R.~A.},
  \bibinfo{author}{{Kurtz}, S.~E.} et~al. (\bibinfo{year}{2023}).
\newblock \bibinfo{title}{{Nature of continuum emission in the source of the
  water maser super-flare G25.65+1.04}}.
\newblock {\it \bibinfo{journal}{Astronomy \& Astrophysics}\/},  {\it
  \bibinfo{volume}{673}\/}, \bibinfo{pages}{A60}.
  \DOIprefix\doi{10.1051/0004-6361/202346023}.

\bibitem[{{Bayandina} et~al.(2019){Bayandina}, {Burns}, {Kurtz},
  {Shakhvorostova} \& {Val'tts}}]{Bayandina2019}
\bibinfo{author}{{Bayandina}, O.~S.}, \bibinfo{author}{{Burns}, R.~A.},
  \bibinfo{author}{{Kurtz}, S.~E.} et~al. (\bibinfo{year}{2019}).
\newblock \bibinfo{title}{{VLA Overview of the Bursting H2O Maser Source
  G25.65+1.05}}.
\newblock {\it \bibinfo{journal}{The Astrophysical Journal}\/},  {\it
  \bibinfo{volume}{884}\/}\bibinfo{issue}{(2)}, \bibinfo{pages}{140}.
  \DOIprefix\doi{10.3847/1538-4357/ab3fa4}.
  \href{http://arxiv.org/abs/1812.11353}{\tt arXiv:1812.11353}.

\bibitem[{{Bayandina} et~al.(2020){Bayandina}, {Shakhvorostova}, {Alakoz},
  {Burns}, {Kurtz} \& {Val'tts}}]{Bayandina2020}
\bibinfo{author}{{Bayandina}, O.~S.}, \bibinfo{author}{{Shakhvorostova},
  N.~N.}, \bibinfo{author}{{Alakoz}, A.~V.} et~al. (\bibinfo{year}{2020}).
\newblock \bibinfo{title}{{RadioAstron reveals super-compact structures in the
  bursting H$_{2}$O maser source G25.65+1.05}}.
\newblock {\it \bibinfo{journal}{Advances in Space Research}\/},  {\it
  \bibinfo{volume}{65}\/}\bibinfo{issue}{(2)}, \bibinfo{pages}{763--771}.
  \DOIprefix\doi{10.1016/j.asr.2019.03.011}.

\bibitem[{{Brogan} et~al.(2019){Brogan}, {Hunter}, {Towner}, {McGuire},
  {MacLeod}, {Gurwell}, {Cyganowski}, {Brand}, {Burns}, {Caratti o Garatti},
  {Chen}, {Chibueze}, {Hirano}, {Hirota}, {Kim}, {Kramer}, {Linz}, {Menten},
  {Remijan}, {Sanna}, {Sobolev}, {Sridharan}, {Stecklum}, {Sugiyama}, {Surcis},
  {Van der Walt}, {Volvach} \& {Volvach}}]{2019ApJ...881L..39B}
\bibinfo{author}{{Brogan}, C.~L.}, \bibinfo{author}{{Hunter}, T.~R.},
  \bibinfo{author}{{Towner}, A.~P.~M.} et~al. (\bibinfo{year}{2019}).
\newblock \bibinfo{title}{{Sub-arcsecond (Sub)millimeter Imaging of the Massive
  Protocluster G358.93-0.03: Discovery of 14 New Methanol Maser Lines
  Associated with a Hot Core}}.
\newblock {\it \bibinfo{journal}{The Astrophysical Journal Letters}\/},  {\it
  \bibinfo{volume}{881}\/}\bibinfo{issue}{(2)}, \bibinfo{pages}{L39}.
  \DOIprefix\doi{10.3847/2041-8213/ab2f8a}.
  \href{http://arxiv.org/abs/1907.02470}{\tt arXiv:1907.02470}.

\bibitem[{Burns et~al.(2018)Burns, Bayandina, Orosz, Olech, Immer, Blanchard,
  Marcote, van Langevelde, Hirota, Kim, Valtts, Shakhvorostova, Rudnitskii,
  Volvach, Volvach, MacLeod, Chibueze, Surcis, Kramer, Baan, Brogan, Hunter \&
  Kurtz}]{burns2018}
\bibinfo{author}{Burns, R.~A.}, \bibinfo{author}{Bayandina, O.},
  \bibinfo{author}{Orosz, G.} et~al. (\bibinfo{year}{2018}).
\newblock \bibinfo{title}{Multi-epoch vlbi of a double maser super burst}.
\newblock \URLprefix \url{https://arxiv.org/abs/1812.09454}.
  \href{http://arxiv.org/abs/1812.09454}{\tt arXiv:1812.09454}.

\bibitem[{{Burns} et~al.(2020{\natexlab{a}}){Burns}, {Orosz}, {Bayandina},
  {Surcis}, {Olech}, {MacLeod}, {Volvach}, {Rudnitskii}, {Hirota}, {Immer},
  {Blanchard}, {Marcote}, {van Langevelde}, {Chibueze}, {Sugiyama}, {Kim},
  {Val`tts}, {Shakhvorostova}, {Kramer}, {Baan}, {Brogan}, {Hunter}, {Kurtz},
  {Sobolev}, {Brand} \& {Volvach}}]{Burns2020}
\bibinfo{author}{{Burns}, R.~A.}, \bibinfo{author}{{Orosz}, G.},
  \bibinfo{author}{{Bayandina}, O.} et~al.
  (\bibinfo{year}{2020}{\natexlab{a}}).
\newblock \bibinfo{title}{{VLBI observations of the G25.65+1.05 water maser
  superburst}}.
\newblock {\it \bibinfo{journal}{Monthly Notices of the Royal Astronomical
  Society}\/},  {\it \bibinfo{volume}{491}\/}\bibinfo{issue}{(3)},
  \bibinfo{pages}{4069--4075}. \DOIprefix\doi{10.1093/mnras/stz3172}.
  \href{http://arxiv.org/abs/1911.12634}{\tt arXiv:1911.12634}.

\bibitem[{{Burns} et~al.(2020{\natexlab{b}}){Burns}, {Sugiyama}, {Hirota},
  {Kim}, {Sobolev}, {Stecklum}, {MacLeod}, {Yonekura}, {Olech}, {Orosz},
  {Ellingsen}, {Hyland}, {Caratti o Garatti}, {Brogan}, {Hunter}, {Phillips},
  {van den Heever}, {Eisl{\"o}ffel}, {Linz}, {Surcis}, {Chibueze}, {Baan} \&
  {Kramer}}]{Burns20}
\bibinfo{author}{{Burns}, R.~A.}, \bibinfo{author}{{Sugiyama}, K.},
  \bibinfo{author}{{Hirota}, T.} et~al. (\bibinfo{year}{2020}{\natexlab{b}}).
\newblock \bibinfo{title}{{A heatwave of accretion energy traced by masers in
  the G358-MM1 high-mass protostar}}.
\newblock {\it \bibinfo{journal}{Nature Astronomy}\/},  {\it
  \bibinfo{volume}{4}\/}, \bibinfo{pages}{506--510}.
  \DOIprefix\doi{10.1038/s41550-019-0989-3}.

\bibitem[{{Caratti o Garatti} et~al.(2017){Caratti o Garatti}, {Stecklum},
  {Garcia Lopez}, {Eisl{\"o}ffel}, {Ray}, {Sanna}, {Cesaroni}, {Walmsley},
  {Oudmaijer}, {de Wit}, {Moscadelli}, {Greiner}, {Krabbe}, {Fischer}, {Klein}
  \& {Iba{\~n}ez}}]{2017NatPh..13..276C}
\bibinfo{author}{{Caratti o Garatti}, A.}, \bibinfo{author}{{Stecklum}, B.},
  \bibinfo{author}{{Garcia Lopez}, R.} et~al. (\bibinfo{year}{2017}).
\newblock \bibinfo{title}{{Disk-mediated accretion burst in a high-mass young
  stellar object}}.
\newblock {\it \bibinfo{journal}{Nature Physics}\/},  {\it
  \bibinfo{volume}{13}\/}\bibinfo{issue}{(3)}, \bibinfo{pages}{276--279}.
  \DOIprefix\doi{10.1038/nphys3942}. \href{http://arxiv.org/abs/1704.02628}{\tt
  arXiv:1704.02628}.

\bibitem[{{Colom} et~al.(2019){Colom}, {Ashimbaeva}, {Lekht}, {Pashchenko},
  {Rudnitskii}, {Krasnov} \& {Tolmachev}}]{2019ARep...63..814C}
\bibinfo{author}{{Colom}, P.}, \bibinfo{author}{{Ashimbaeva}, N.~T.},
  \bibinfo{author}{{Lekht}, E.~E.} et~al. (\bibinfo{year}{2019}).
\newblock \bibinfo{title}{{Observations of Maser Emission in the Star-Forming
  Region G43.8-0.1. II. H$_{2}$O Maser Emission at 1.35 cm}}.
\newblock {\it \bibinfo{journal}{Astronomy Reports}\/},  {\it
  \bibinfo{volume}{63}\/}\bibinfo{issue}{(10)}, \bibinfo{pages}{814--829}.
  \DOIprefix\doi{10.1134/S0004629919100049}.

\bibitem[{Davis et~al.(1985)Davis, Herring, Shapiro, Rogers \&
  Elgered}]{Davis1985}
\bibinfo{author}{Davis, J.~L.}, \bibinfo{author}{Herring, T.~A.},
  \bibinfo{author}{Shapiro, I.~I.} et~al. (\bibinfo{year}{1985}).
\newblock \bibinfo{title}{Geodesy by radio interferometry: Effects of
  atmospheric modeling errors on estimates of baseline length}.
\newblock {\it \bibinfo{journal}{Radio Science}\/},  {\it
  \bibinfo{volume}{20}\/}\bibinfo{issue}{(6)}, \bibinfo{pages}{1593--1607}.
  \DOIprefix\doi{https://doi.org/10.1029/RS020i006p01593}.

\bibitem[{{Elitzur} et~al.(1989){Elitzur}, {Hollenbach} \&
  {McKee}}]{1989ApJ...346..983E}
\bibinfo{author}{{Elitzur}, M.}, \bibinfo{author}{{Hollenbach}, D.~J.},  \&
  \bibinfo{author}{{McKee}, C.~F.} (\bibinfo{year}{1989}).
\newblock \bibinfo{title}{{H$_{2}$O Masers in Star-forming Regions}}.
\newblock {\it \bibinfo{journal}{Astrophysical Journal}\/},  {\it
  \bibinfo{volume}{346}\/}, \bibinfo{pages}{983}.
  \DOIprefix\doi{10.1086/168080}.

\bibitem[{{Hirota} et~al.(2011){Hirota}, {Tsuboi}, {Fujisawa}, {Honma},
  {Kawaguchi}, {Kim}, {Kobayashi}, {Imai}, {Omodaka}, {Shibata}, {Shimoikura}
  \& {Yonekura}}]{2011ApJ...739L..59H}
\bibinfo{author}{{Hirota}, T.}, \bibinfo{author}{{Tsuboi}, M.},
  \bibinfo{author}{{Fujisawa}, K.} et~al. (\bibinfo{year}{2011}).
\newblock \bibinfo{title}{{Identification of Bursting Water Maser Features in
  Orion KL}}.
\newblock {\it \bibinfo{journal}{The Astrophysical Journal Letters}\/},  {\it
  \bibinfo{volume}{739}\/}\bibinfo{issue}{(2)}, \bibinfo{pages}{L59}.
  \DOIprefix\doi{10.1088/2041-8205/739/2/L59}.
  \href{http://arxiv.org/abs/1108.3889}{\tt arXiv:1108.3889}.

\bibitem[{{Hirota} et~al.(2014){Hirota}, {Tsuboi}, {Kurono}, {Fujisawa},
  {Honma}, {Kim}, {Imai} \& {Yonekura}}]{2014PASJ...66..106H}
\bibinfo{author}{{Hirota}, T.}, \bibinfo{author}{{Tsuboi}, M.},
  \bibinfo{author}{{Kurono}, Y.} et~al. (\bibinfo{year}{2014}).
\newblock \bibinfo{title}{{VERA and ALMA observations of the H$_{2}$O
  supermaser burst in Orion KL}}.
\newblock {\it \bibinfo{journal}{Publications of the Astronomical Society of
  Japan}\/},  {\it \bibinfo{volume}{66}\/}\bibinfo{issue}{(6)},
  \bibinfo{pages}{106}. \DOIprefix\doi{10.1093/pasj/psu110}.
  \href{http://arxiv.org/abs/1407.2757}{\tt arXiv:1407.2757}.

\bibitem[{{Hollenbach} et~al.(2013){Hollenbach}, {Elitzur} \&
  {McKee}}]{2013ApJ...773...70H}
\bibinfo{author}{{Hollenbach}, D.}, \bibinfo{author}{{Elitzur}, M.},  \&
  \bibinfo{author}{{McKee}, C.~F.} (\bibinfo{year}{2013}).
\newblock \bibinfo{title}{{Interstellar H$_{2}$O Masers from J Shocks}}.
\newblock {\it \bibinfo{journal}{Astrophysical Journal}\/},  {\it
  \bibinfo{volume}{773}\/}\bibinfo{issue}{(1)}, \bibinfo{pages}{70}.
  \DOIprefix\doi{10.1088/0004-637X/773/1/70}.
  \href{http://arxiv.org/abs/1306.5276}{\tt arXiv:1306.5276}.

\bibitem[{{Kardashev} et~al.(2017){Kardashev}, {Alakoz}, {Andrianov},
  {Artyukhov}, {Baan}, {Babyshkin}, {Bartel}, {Bayandina}, {Val'tts},
  {Voitsik}, {Vorobyov}, {Gwinn}, {Gomez}, {Giovannini}, {Jauncey}, {Johnson},
  {Imai}, {Kovalev}, {Kurtz}, {Lisakov}, {Lobanov}, {Molodtsov}, {Novikov},
  {Pogodin}, {Popov}, {Privesenzev}, {Rudnitski}, {Rudnitski}, {Savolainen},
  {Smirnova}, {Sobolev}, {Soglasnov}, {Sokolovsky}, {Filippova}, {Khartov},
  {Churikova}, {Shirshakov}, {Shishov} \& {Edwards}}]{Kardashev2017}
\bibinfo{author}{{Kardashev}, N.~S.}, \bibinfo{author}{{Alakoz}, A.~V.},
  \bibinfo{author}{{Andrianov}, A.~S.} et~al. (\bibinfo{year}{2017}).
\newblock \bibinfo{title}{{RadioAstron Science Program Five Years after Launch:
  Main Science Results}}.
\newblock {\it \bibinfo{journal}{Solar System Research}\/},  {\it
  \bibinfo{volume}{51}\/}\bibinfo{issue}{(7)}, \bibinfo{pages}{535--554}.
  \DOIprefix\doi{10.1134/S0038094617070085}.

\bibitem[{{Kardashev} et~al.(2013){Kardashev}, {Khartov}, {Abramov}, {Avdeev},
  {Alakoz}, {Aleksandrov}, {Ananthakrishnan}, {Andreyanov}, {Andrianov},
  {Antonov}, {Artyukhov}, {Arkhipov}, {Baan}, {Babakin}, {Babyshkin},
  {Bartel'}, {Belousov}, {Belyaev}, {Berulis}, {Burke}, {Biryukov}, {Bubnov},
  {Burgin}, {Busca}, {Bykadorov}, {Bychkova}, {Vasil'kov}, {Wellington},
  {Vinogradov}, {Wietfeldt}, {Voitsik}, {Gvamichava}, {Girin}, {Gurvits},
  {Dagkesamanskii}, {D'Addario}, {Giovannini}, {Jauncey}, {Dewdney}, {D'yakov},
  {Zharov}, {Zhuravlev}, {Zaslavskii}, {Zakhvatkin}, {Zinov'ev}, {Ilinen},
  {Ipatov}, {Kanevskii}, {Knorin}, {Casse}, {Kellermann}, {Kovalev}, {Kovalev},
  {Kovalenko}, {Kogan}, {Komaev}, {Konovalenko}, {Kopelyanskii}, {Korneev},
  {Kostenko}, {Kotik}, {Kreisman}, {Kukushkin}, {Kulishenko}, {Cooper},
  {Kut'kin}, {Cannon}, {Larionov}, {Lisakov}, {Litvinenko}, {Likhachev},
  {Likhacheva}, {Lobanov}, {Logvinenko}, {Langston}, {McCracken}, {Medvedev},
  {Melekhin}, {Menderov}, {Murphy}, {Mizyakina}, {Mozgovoi}, {Nikolaev},
  {Novikov}, {Novikov}, {Oreshko}, {Pavlenko}, {Pashchenko}, {Ponomarev},
  {Popov}, {Pravin-Kumar}, {Preston}, {Pyshnov}, {Rakhimov}, {Rozhkov},
  {Romney}, {Rocha}, {Rudakov}, {R{\"a}is{\"a}nen} et~al.}]{Kardashev2013}
\bibinfo{author}{{Kardashev}, N.~S.}, \bibinfo{author}{{Khartov}, V.~V.},
  \bibinfo{author}{{Abramov}, V.~V.} et~al. (\bibinfo{year}{2013}).
\newblock \bibinfo{title}{{``RadioAstron''-A telescope with a size of 300 000
  km: Main parameters and first observational results}}.
\newblock {\it \bibinfo{journal}{Astronomy Reports}\/},  {\it
  \bibinfo{volume}{57}\/}\bibinfo{issue}{(3)}, \bibinfo{pages}{153--194}.
  \DOIprefix\doi{10.1134/S1063772913030025}.
  \href{http://arxiv.org/abs/1303.5013}{\tt arXiv:1303.5013}.

\bibitem[{{Lekht} et~al.(2018){Lekht}, {Pashchenko}, {Rudnitskii} \&
  {Tolmachev}}]{2018ARep...62..213L}
\bibinfo{author}{{Lekht}, E.~E.}, \bibinfo{author}{{Pashchenko}, M.~I.},
  \bibinfo{author}{{Rudnitskii}, G.~M.} et~al. (\bibinfo{year}{2018}).
\newblock \bibinfo{title}{{Superflares of H$_{2}$O Maser Emission Toward the
  Protostellar Object G25.65+1.05 (IRAS 18316-0602)}}.
\newblock {\it \bibinfo{journal}{Astronomy Reports}\/},  {\it
  \bibinfo{volume}{62}\/}\bibinfo{issue}{(3)}, \bibinfo{pages}{213--224}.
  \DOIprefix\doi{10.1134/S1063772918030071}.
  \href{http://arxiv.org/abs/1709.08197}{\tt arXiv:1709.08197}.

\bibitem[{{Likhachev} et~al.(2017){Likhachev}, {Kostenko}, {Girin},
  {Andrianov}, {Rudnitskiy} \& {Zharov}}]{Likhachev2017}
\bibinfo{author}{{Likhachev}, S.~F.}, \bibinfo{author}{{Kostenko}, V.~I.},
  \bibinfo{author}{{Girin}, I.~A.} et~al. (\bibinfo{year}{2017}).
\newblock \bibinfo{title}{{Software Correlator for Radioastron Mission}}.
\newblock {\it \bibinfo{journal}{Journal of Astronomical Instrumentation}\/},
  {\it \bibinfo{volume}{6}\/}\bibinfo{issue}{(3)},
  \bibinfo{pages}{1750004--131}. \DOIprefix\doi{10.1142/S2251171717500040}.
  \href{http://arxiv.org/abs/1706.06320}{\tt arXiv:1706.06320}.

\bibitem[{Liljeström \& Gwinn(2000)}]{Liljestrom2000}
\bibinfo{author}{Liljeström, T.},  \& \bibinfo{author}{Gwinn, C.~R.}
  (\bibinfo{year}{2000}).
\newblock \bibinfo{title}{Water masers diagnosing postshocked conditions in
  w49n}.
\newblock {\it \bibinfo{journal}{The Astrophysical Journal}\/},  {\it
  \bibinfo{volume}{534}\/}, \bibinfo{pages}{781–800}.
  \DOIprefix\doi{10.1086/308781}.
\newblock \bibinfo{note}{ADS Bibcode: 2000ApJ...534..781L}.

\bibitem[{Murray(1967)}]{Murray1967}
\bibinfo{author}{Murray, F.~W.} (\bibinfo{year}{1967}).
\newblock \bibinfo{title}{On the computation of saturation vapor pressure}.
\newblock {\it \bibinfo{journal}{Journal of Applied Meteorology and
  Climatology}\/},  {\it \bibinfo{volume}{6}\/}\bibinfo{issue}{(1)},
  \bibinfo{pages}{203 -- 204}.
  \DOIprefix\doi{10.1175/1520-0450(1967)006<0203:OTCOSV>2.0.CO;2}.

\bibitem[{{Nesterenok}(2022)}]{2022AstL...48..345N}
\bibinfo{author}{{Nesterenok}, A.~V.} (\bibinfo{year}{2022}).
\newblock \bibinfo{title}{{Collisional Pumping of H\{\}$_{2}$O and
  CH\{\}$_{3}$OH Masers in C-Type Shock Waves}}.
\newblock {\it \bibinfo{journal}{Astronomy Letters}\/},  {\it
  \bibinfo{volume}{48}\/}\bibinfo{issue}{(6)}, \bibinfo{pages}{345--359}.
  \DOIprefix\doi{10.1134/S1063773722060044}.
  \href{http://arxiv.org/abs/2208.00201}{\tt arXiv:2208.00201}.

\bibitem[{Niell(1996)}]{Niell1996}
\bibinfo{author}{Niell, A.~E.} (\bibinfo{year}{1996}).
\newblock \bibinfo{title}{Global mapping functions for the atmosphere delay at
  radio wavelengths}.
\newblock {\it \bibinfo{journal}{Journal of Geophysical Research: Solid
  Earth}\/},  {\it \bibinfo{volume}{101}\/}\bibinfo{issue}{(B2)},
  \bibinfo{pages}{3227--3246}.
  \DOIprefix\doi{https://doi.org/10.1029/95JB03048}.

\bibitem[{Saastamoinen(1972)}]{Saastamoinen1972}
\bibinfo{author}{Saastamoinen, J.} (\bibinfo{year}{1972}).
\newblock \bibinfo{title}{Atmospheric correction for the troposphere and
  stratosphere in radio ranging satellites}.
\newblock In {\it \bibinfo{booktitle}{The Use of Artificial Satellites for
  Geodesy}\/} (pp. \bibinfo{pages}{247--251}).
\newblock \bibinfo{publisher}{American Geophysical Union (AGU)}.
\newblock \DOIprefix\doi{https://doi.org/10.1029/GM015p0247}.

\bibitem[{{Shakhvorostova} et~al.(2018){Shakhvorostova}, {Vol'vach},
  {Vol'vach}, {Dmitrotsa}, {Bayandina}, {Val'tts}, {Alakoz}, {Ashimbaeva} \&
  {Rudnitskii}}]{2018ARep...62..584S}
\bibinfo{author}{{Shakhvorostova}, N.~N.}, \bibinfo{author}{{Vol'vach}, L.~N.},
  \bibinfo{author}{{Vol'vach}, A.~E.} et~al. (\bibinfo{year}{2018}).
\newblock \bibinfo{title}{{Search for H$_{2}$O Maser Flares in Regions of
  Formation of Massive Stars}}.
\newblock {\it \bibinfo{journal}{Astronomy Reports}\/},  {\it
  \bibinfo{volume}{62}\/}\bibinfo{issue}{(9)}, \bibinfo{pages}{584--608}.
  \DOIprefix\doi{10.1134/S1063772918090081}.

\bibitem[{{Shimoikura} et~al.(2005){Shimoikura}, {Kobayashi}, {Omodaka},
  {Diamond}, {Matveyenko} \& {Fujisawa}}]{2005ApJ...634..459S}
\bibinfo{author}{{Shimoikura}, T.}, \bibinfo{author}{{Kobayashi}, H.},
  \bibinfo{author}{{Omodaka}, T.} et~al. (\bibinfo{year}{2005}).
\newblock \bibinfo{title}{{VLBA Observations of a Bursting Water Maser in Orion
  KL}}.
\newblock {\it \bibinfo{journal}{The Astrophysical Journal}\/},  {\it
  \bibinfo{volume}{634}\/}\bibinfo{issue}{(1)}, \bibinfo{pages}{459--467}.
  \DOIprefix\doi{10.1086/432865}.

\bibitem[{{Sobolev} et~al.(2018){Sobolev}, {Moran}, {Gray}, {Alakoz}, {Imai},
  {Baan}, {Tolmachev}, {Samodurov} \& {Ladeyshchikov}}]{2018ApJ...856...60S}
\bibinfo{author}{{Sobolev}, A.~M.}, \bibinfo{author}{{Moran}, J.~M.},
  \bibinfo{author}{{Gray}, M.~D.} et~al. (\bibinfo{year}{2018}).
\newblock \bibinfo{title}{{Sun-sized Water Vapor Masers in Cepheus A}}.
\newblock {\it \bibinfo{journal}{Astrophysical Journal}\/},  {\it
  \bibinfo{volume}{856}\/}\bibinfo{issue}{(1)}, \bibinfo{pages}{60}.
  \DOIprefix\doi{10.3847/1538-4357/aab096}.
  \href{http://arxiv.org/abs/1802.06756}{\tt arXiv:1802.06756}.

\bibitem[{{Sugiyama} et~al.(2019){Sugiyama}, {Saito}, {Yonekura} \&
  {Momose}}]{Sugiyama19}
\bibinfo{author}{{Sugiyama}, K.}, \bibinfo{author}{{Saito}, Y.},
  \bibinfo{author}{{Yonekura}, Y.} et~al. (\bibinfo{year}{2019}).
\newblock \bibinfo{title}{{Bursting activity of the 6.668-GHz CH3OH maser
  detected in G 358.93-00.03 using the Hitachi 32-m}}.
\newblock {\it \bibinfo{journal}{The Astronomer's Telegram}\/},  {\it
  \bibinfo{volume}{12446}\/}, \bibinfo{pages}{1}.

\bibitem[{{Sunada} et~al.(2007){Sunada}, {Nakazato}, {Ikeda}, {Hongo},
  {Kitamura} \& {Yang}}]{2007PASJ...59.1185S}
\bibinfo{author}{{Sunada}, K.}, \bibinfo{author}{{Nakazato}, T.},
  \bibinfo{author}{{Ikeda}, N.} et~al. (\bibinfo{year}{2007}).
\newblock \bibinfo{title}{{Water Maser and Ammonia Survey toward IRAS Sources
  in the Galaxy I. H$_{2}$O Maser Data}}.
\newblock {\it \bibinfo{journal}{PASJ}\/},  {\it \bibinfo{volume}{59}\/},
  \bibinfo{pages}{1185}. \DOIprefix\doi{10.1093/pasj/59.6.1185}.

\bibitem[{{Volvach} et~al.(2017{\natexlab{a}}){Volvach}, {Volvach}, {MacLeod},
  {Bayandina}, {Shakhvorostova} \& {Val'tts}}]{2017ATel10853....1V}
\bibinfo{author}{{Volvach}, A.~E.}, \bibinfo{author}{{Volvach}, L.~N.},
  \bibinfo{author}{{MacLeod}, G.} et~al. (\bibinfo{year}{2017}{\natexlab{a}}).
\newblock \bibinfo{title}{{Detection of a Second Bright H2O Maser Burst from
  G25.65+1.05 at the Simeiz RT-22 and HartRAO Radio Telescopes}}.
\newblock {\it \bibinfo{journal}{The Astronomer's Telegram}\/},  {\it
  \bibinfo{volume}{10853}\/}, \bibinfo{pages}{1}.

\bibitem[{{Volvach} et~al.(2017{\natexlab{b}}){Volvach}, {Volvach}, {MacLeod},
  {Lekht}, {Rudnitskij} \& {Tolmachev}}]{2017ATel10728....1V}
\bibinfo{author}{{Volvach}, A.~E.}, \bibinfo{author}{{Volvach}, L.~N.},
  \bibinfo{author}{{MacLeod}, G.} et~al. (\bibinfo{year}{2017}{\natexlab{b}}).
\newblock \bibinfo{title}{{Detection of a Bright H2O Maser Burst from
  G25.65+1.05 at the Simeiz Radio Telescope RT-22}}.
\newblock {\it \bibinfo{journal}{The Astronomer's Telegram}\/},  {\it
  \bibinfo{volume}{10728}\/}, \bibinfo{pages}{1}.

\bibitem[{{Vol'vach} et~al.(2019){Vol'vach}, {Vol'vach}, {Larionov}, {MacLeod},
  {van den Heever}, {Wolak}, {Olech}, {Ipatov}, {Ivanov}, {Mikhailov},
  {Mel'nikov}, {Menten}, {Belloche}, {Weiss}, {Mazumdar} \&
  {Schuller}}]{2019ARep...63...49V}
\bibinfo{author}{{Vol'vach}, L.~N.}, \bibinfo{author}{{Vol'vach}, A.~E.},
  \bibinfo{author}{{Larionov}, M.~G.} et~al. (\bibinfo{year}{2019}).
\newblock \bibinfo{title}{{A Giant Water Maser Flare in the Galactic Source
  IRAS 18316-0602}}.
\newblock {\it \bibinfo{journal}{Astronomy Reports}\/},  {\it
  \bibinfo{volume}{63}\/}\bibinfo{issue}{(1)}, \bibinfo{pages}{49--65}.
  \DOIprefix\doi{10.1134/S1063772919010062}.

\end{thebibliography}

\end{document}